\documentclass[10pt]{article}
\usepackage{epsfig,amsmath,amssymb,amsfonts,bm,subfigure,setspace}
\begin{document}

\newcommand{\be}{\begin{equation}}
\newcommand{\ee}{\end{equation}}
\newcommand{\bea}{\begin{eqnarray}}
\newcommand{\eea}{\end{eqnarray}}
\newcommand{\nn}{\nonumber}
\newcommand{\ba}{\bea \begin{array}}
\newcommand{\ea}{\end{array} \eea}
\renewcommand{\(}{\left(}
\renewcommand{\)}{\right)}
\renewcommand{\[}{\left[}
\renewcommand{\]}{\right]}
\newcommand{\bc}{\begin{center}}
\newcommand{\ec}{\end{center}}
\newcommand{\pa}{\partial}

\newcommand{\mb}[1]{ \mbox{\boldmath$#1$} }
\newcommand{\ds}{\displaystyle}
\newcommand{\beq}{\begin{eqnarray}}
\newcommand{\eeq}{\end{eqnarray}}
\newcommand{\beqq}{\begin{eqnarray*}}
\newcommand{\eeqq}{\end{eqnarray*}}
\newcommand{\p}{\partial}
\newcommand{\g}{\gamma}
\newcommand{\eps}{\varepsilon}
\newcommand{\x}{\mbox{\boldmath$x$}}
\newcommand{\e}{\mbox{\boldmath$e$}}
\newcommand{\n}{\mbox{\boldmath$n$}}
\newcommand{\J}{\mbox{\boldmath$J$}}
\newcommand{\y}{\mbox{\boldmath$y$}}
\newcommand{\w}{\mbox{\boldmath$w$}}
\newcommand{\z}{\mbox{\boldmath$z$}}
\font\bb=msbm10 at 12pt
\def\rR{\hbox{\bb R}}
\def\rN{\hbox{\bb N}}
\def\rQ{\hbox{\bb Q}}
\def\rZ{\hbox{\bb Z}}

\title{The Narrow Escape problem in a flat cylindrical microdomain with application to diffusion in the synaptic cleft}

\author{J\"urgen Reingruber ~and David
Holcman \thanks{Department of Computational Biology (IBENS) and Mathematics, Ecole Normale
Sup{\'e}rieure, 46 rue d'Ulm 75005 Paris, France.}}
\date{}

\maketitle

\begin{abstract}
The mean first passage time (MFPT) for a Brownian particle to reach
a small target in cellular microdomains is a key parameter for
chemical activation. Although asymptotic estimations of the MFPT are
available for various geometries, these formula cannot be applied to
degenerated structures where one dimension of is much smaller
compared to the others. Here we study the narrow escape time (NET)
problem for a Brownian particle to reach a small target located on
the surface of a flat cylinder, where the cylinder height is
comparable to the target size, and much smaller than the cylinder
radius. When the cylinder is sealed, we estimate the MFPT
for a Brownian particle to hit a small disk located centrally
on the lower surface. For a laterally open cylinder, we estimate the
conditional probability and the conditional MFPT to reach the small
disk before exiting through the lateral opening. We apply our
results to diffusion in the narrow synaptic cleft, and compute the
fraction and the mean time for neurotransmitters to find their
specific receptors located on the postsynaptic terminal. Finally, we
confirm our formulas with Brownian simulations.
\end{abstract}

%
\section{Introduction}
%
{The problem of computing the mean first passage time (MFPT) for a Brownian particle to reach a small
target located on a surface of a microdomain, also referred to as the Narrow
Escape Time (NET) \cite{Ward1,HolcmanSchuss2004_StatPhys}, is
ubiquitous in biophysics and cellular biology because it
corresponds to determining the forward binding rate of chemical
reactions
\cite{ZwanzigPNAS1990,BergPurcell1977,WilemskiFixman1973,Grigorievetal2002}.
Applications of the NET ranges from quantitative analysis for the
resident time of receptors in the postsynaptic density
\cite{HolcmanSchuss2004_StatPhys,HolcmanPNAS2007,TafliaHolcman2007,Holcman_Review2009}, a fundamental microdomain
associated to synaptic transmission and plasticity
\cite{Nicoll2007}, to scaling laws in physics \cite{Klafteretal_Nature2007}, early
steps of viral infection
\cite{Holcman_ViralDNATrafficking_JSP2007,HolcmanLagache_PRE2008,LagacheDautyHolcman_CurrOp2009},
or the hydrolysis rate of activated phosphodiesterase in rod
photoreceptors \cite{ReingruberHolcman_JCP2008,ReingruberHolcman_PRE2009}.

Recent analytical approaches lead to asymptotic formula for the NET
in a confined geometry \cite{Ward1,Ward3,SSH_PRE2008,HolcmanetalNE3}. For example, in a three
dimensional domain of volume $V$ with isoperimetric ratio of order
1, and with no bottlenecks, the overall NET to an absorbing circular
hole of (dimensionless) radius $a$ centered at $\x_s$ on the surface
is \cite{SSH_PRE2008}
\beq \label{aeps}
\tau = \frac{|V|}{4aD}\(1+\frac{L(\x_s)+N(\x_s)}{2\pi}a\ln a +O(a\ln a))\)^{-1} \,,
\eeq
where $D$ is the diffusion constant, $L(\x_s)$ and $N(\x_s)$ are the principal curvatures at $x_s$.
In the case of a sphere, a precise asymptotic expression with the
first three terms was recently obtained in \cite{WardCheviakov_Siam2010},
where the $O(1)$ term depends on the regular part of the
Green's function. The NET computations were further generalized to the case of several holes
\cite{HolcmanSchuss_PhysLettA2008,WardPillay_Siam2010,WardCheviakov_Siam2010},
and to stochastic dynamics with a potential well
\cite{SingerSchuss_PRE2006,TafliaHolcman2007}.

However, the NET formula (\ref{aeps}) cannot be directly applied to degenerated microdomains
where one dimension is much smaller than the others. This is for
example the case for the synaptic cleft separating pre- and
post-synaptic neuronal terminals (Fig.~\ref{Domain_cylinder}a), which can be
approximated as a flat cylinder with height much smaller compared to its width
\cite{Harris_Spines_RevNeurosc2008}. Furthermore, in retinal rod
photoreceptors sustaining night vision, the outer segment
contains thousands of piled flat cylinders that define the photoresponse and
the fidelity of the vision under dim light conditions \cite{RevRiekeBaylor1998,HolcmanKorenbrot2004,ReingruberHolcman_JCP2008}.

The goal of this paper is to extend the NET analysis to degenerate
domains. More specifically, we study the NET of a Brownian
particle in a flat cylinder, where the cylinder height $h$ is much smaller
compared to the cylinder radius $R$ ($h \ll R$), with a small circular hole
of radius $a$ centered on the bottom cylinder surface (Fig.~\ref{Domain_cylinder}b).
In the first part, we will analyze
the NET to exit the cylindrical domain when the boundary is
reflecting everywhere except at the small hole, where it is absorbing. Due to the radial symmetry, the
solution of the the mixed boundary value problem can be expanded in
terms of Bessel functions. For a flat cylinder with $h \ll R$ and $R \gg a$,
we find that the NET is given by
\bea \label{newfintro}
\tau \approx \frac{|V|}{aD}\frac{a_0\(\frac{h}{a}\)}{\sqrt 2}  + \frac{R^2}{8D} \(4\ln\(\frac{R}{a}\) -3 \)\,,
\eea
where the function ${a_0\(\frac{h}{a}\)}/{\sqrt 2} \in [0.07,0.25]$ is depicted in Fig.~\ref{fig_a0}a. Although we derive (\ref{newfintro}) for $h \ll R$, we expect that it remains a valid approximation until $h\sim R$, in which case $a_0(\frac{h}{a})\sim \frac{1}{4}$ and the leading order terms in (\ref{newfintro}) and (\ref{aeps}) coincide. We note that the log-contribution in (\ref{aeps}) comes from the local property of the boundary at the hole, whereas in (\ref{newfintro}) it originates from the degenerated geometry.

{In the second part of the paper, we study a cylinder that is open at lateral boundary, and we present asymptotic estimates for the conditional probability $p$ and the conditional mean time $\tau_c$ that a Brownian particle
reaches the small hole before leaving the domain through the lateral boundary. For example, for a flat cylinder with $h\sim a$ and $R\gg a$, when the particle starts at the upper boundary at position $(r=0,z=h)$ opposite to the small hole, the conditional probability $p(0,h)$ and the conditional mean time $\tau_c(0,h)$ are ((\ref{approxtauuppersurf}), (\ref{p(x,y)largealpha_tau}) and (\ref{approxtauctau}))
\bea
\begin{array}{rcl}
\ds  \tau(0,h) &\approx & \ds{ \frac{|V|}{aD} \frac{a_0(h/a)}{I_0(\frac{\pi a}{2h})}}, \\ \\
\ds{p(0,h)} &\approx&  \ds{ 1 -\frac{2 D}{R^2\ln (R/a)} \tau(0,h)},\\ \\
\ds \tau_c(0,h) &\approx &  \ds{ \frac{\ds{1 - \frac{\sqrt 2 D}{R^2 \ln (R/a)}} \ds{I_0(\frac{\pi a}{2 h})} \tau(0,h)}{ \ds{1 -\ \frac{2 D}{R^2 \ln(R/a)} \tau(0,h)}}}\ds{ \frac{ \tau(0,h)}{2 (\ln(R/a))^2}},
\end{array}
\eea
where $\alpha=R/a\gg1$, $\beta=h/a \sim 1$ and $\tau(0,h)$ is the mean time to reach the small hole when the cylinder is closed. These asymptotic expressions can be applied to study diffusion in the synaptic cleft, where synaptic transmission depends on neurotransmitters that are released at the presynaptic terminal from vesicles and activate receptors located on the opposite post-synaptic neuron (Fig.~\ref{Domain_cylinder}). The transmission efficiency depends crucially on the conditional probability for a diffusing neurotransmitter to hit
a functional receptor before leaving the synaptic cleft.

%
\section{Mean time to find a small target in a bounded cylindrical
compartment} \label{section_tau_cylinder}
We shall now present our analysis to estimate the MFPT $\tau(\x)$
for a Brownian molecule, initially located at position
$\x=(x_1,x_2,z)$, to escape a cylinder of radius $R$ and height $h$
(Fig.\ref{Domain_cylinder}) through a small circular hole of radius
$a$ located centrally on the lower surface at $z=0$. The cylindrical
surface is reflecting, except for the small hole where it is
absorbing. Due to the radial symmetry, the MFPT is a function of
the radius $r=\sqrt{x_1^2 +x_2^2}$ and the height $z$. Using the
small hole radius $a$, we define the dimensionless parameters and
variables
\bea \label{defscaling}
x=\frac{r}{a}\,, \quad y=\frac{z}{a}\,, \quad \alpha=\frac{R}{a}\,, \quad  \beta=\frac{h}{a}\,,
\quad |\Omega| =\frac{|V|}{a^3}=\pi\beta \alpha^2\,,\nn
\eea
and the scaled MFPT
\bea \label{defscalingTau}
\hat \tau(x,y)=\frac{a D}{|V|} \tau(r,z) = \frac{D}{\pi R^2 \beta}  \tau(r,z) \,,
\eea
which is a solution of \cite{BookSchuss}
\bea
\begin{array}{rcl}
\ds\( \frac{1}{x} \frac{\p}{\p x} x\frac{\p}{\p x} +
\frac{\p^2}{\p y^2}\) \hat \tau (x,y) &=& \ds -\frac{1}{|\Omega|} \,, \quad x \in \Omega \label{scaledEqtau_2}\\
\ds \hat \tau(x,y)&=& 0\,, \quad y=0\,, \, x < 1 \\
\ds \frac{\partial}{\partial y}\hat \tau (x,y)&=&0 \,, \quad  y=0\,, \, 1 < x < \alpha  \\
 \ds \frac{\partial}{\partial y}\hat \tau (x,y)\Big |_{y=\beta}=0\,,
&& \ds \frac{\partial}{\partial x}\hat \tau (x,y)\Big |_{x=\alpha} =0 \,.
\end{array}
\eea
Our goal is to obtain a solution for (\ref{scaledEqtau_2}) and to clarify its dependency on the parameters $\alpha$ and $\beta$. To study the shape of the boundary layer, we note that
(\ref{scaledEqtau_2}) corresponds to a heat equation
where the total amount of heat produced in $\Omega$ is one,
independent of $\alpha$ and $\beta$. Furthermore, because the scaled
radius of the hole through which the heat dissipates is one, it follows that $\hat \tau
(x,y)$ has a finite asymptotic limit in the neighborhood of the hole for $\alpha\to \infty $ and $\beta \to \infty$.
\begin{figure}[ht!]
  \begin{center}
   \subfigure[]{\includegraphics[scale=0.2]{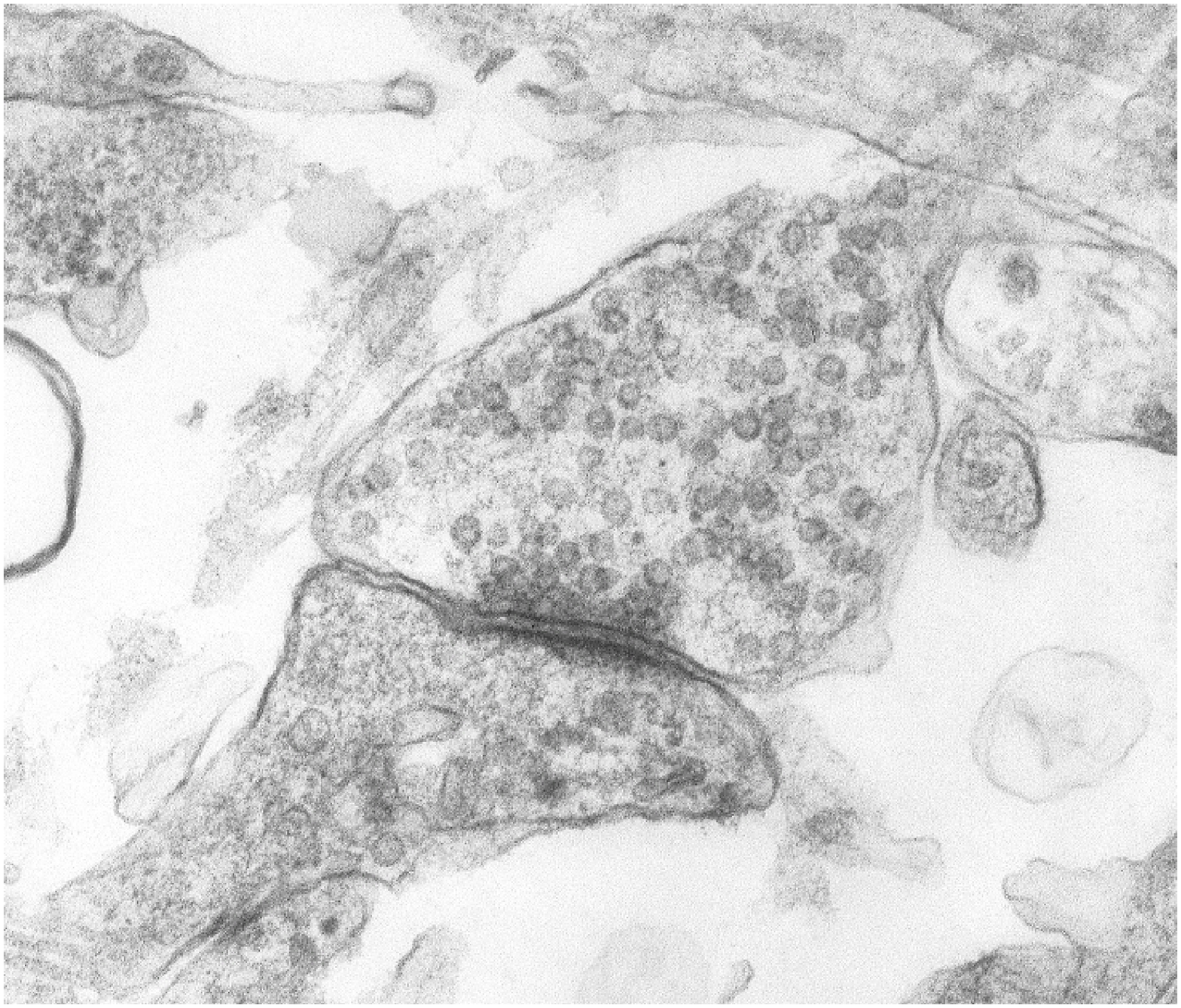}}
   \hspace{1cm}
    \subfigure[]{\includegraphics[scale=0.5]{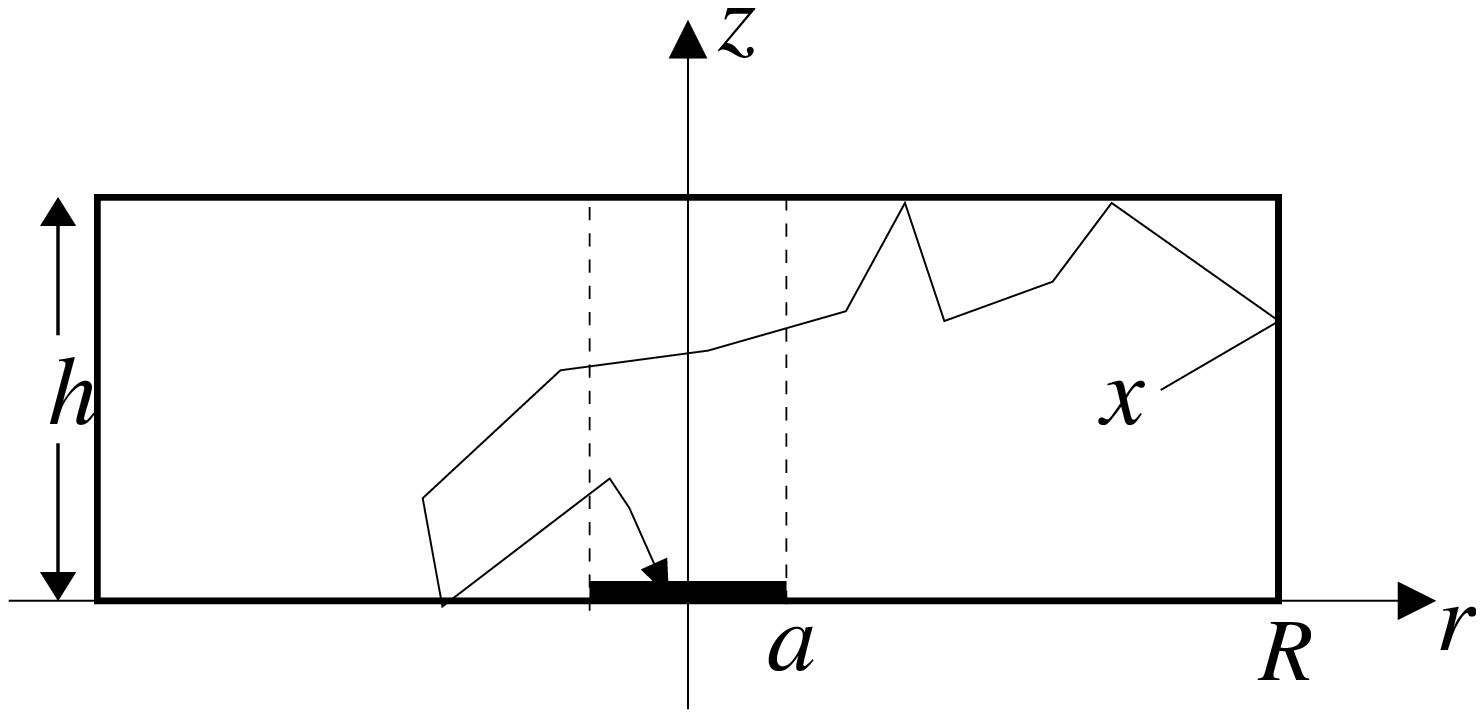}}
    \caption{(a) EM picture of a synapse showing a synaptic cleft and the two pre and post-synaptic terminals. (b) { Schematic representation
of a Brownian trajectory in a cylinder of height $h$ and radius $R$
with reflecting boundaries, except at the small absorbing disk of
radius $a\ll R$, located centrally on the lower surface. We are interested in the case of a flat cylinder with $ R \gg a$ and $h \sim a$.} }
\label{Domain_cylinder}
  \end{center}
\end{figure}

%
\subsection{Equation for the scaled MFPT $\hat \tau(x,y)$}
%
To derive the solution $\hat\tau(x,y)$ of (\ref{scaledEqtau_2}),
we consider the two domains
\beq
\Omega_i=\{ x|x < 1\} \hbox{ and }
\Omega_o=\{ x |1 < x < \alpha\}, \nn
\eeq
and obtain for $\hat\tau(x,y)$ the representation
\bea\label{compositeTau}
\hat \tau(x,y)=
\left\{
\begin{array}{l}
\displaystyle  \hat \tau_i(x,y) \,, \quad (x,y) \in \Omega_i \\ \\
\displaystyle  \hat \tau_o(x,y) \,, \quad  (x,y) \in \Omega_o\,.
\end{array}\right.
\eea
To ensure that $\hat \tau(x,y)$ is a solution of
(\ref{scaledEqtau_2}) in $\Omega$, $\hat \tau(x,y)$ and the flux
$\frac{\p}{\p x}\hat\tau(x,y)$ have to be continuous at $x=1$,
leading to the conditions
\bea
\begin{array}{lll}
\ds \hat \tau_i(1,y) &=& \ds \hat \tau_o(1,y)\,, \quad 0< y < \beta \label{condtau1} \\ \\
\ds \frac{\p \hat \tau_i(x,y)}{\p x}\Big|_{x=1} &=& \ds \frac{\p \hat \tau_o(x,y)}{\p x} \Big|_{x=1}\,, \quad 0< y < \beta \,. \label{condtau2}
\end{array}
\eea
Using a separation of variable method, we expand $\tau_i(x,y)$ and
$\tau_o(x,y)$ in series
\bea
\tau_i(x,y) &=& \sum_{n=0}^\infty b_n q_n(x) v_n(y) + w_i(x,y) \label{deftaui}\\
\tau_o(x,y) &=& \sum_{n=0}^\infty a_n p_n(x) u_n(y) +w_o(x,y) \label{deftauo}
\eea
where
\bea
\begin{array}{c}
\ds k_n=\frac{n\pi}{\beta}\,,\quad l_n=\frac{(n+\frac{1}{2})\pi}{\beta}\,, \\
\ds u_0=\frac{1}{\sqrt 2}\,, \quad  u_n(y)= \cos\(k_n y \) \quad (n\ge 1)\,, \quad  v_n(y) = \sin(l_n y )  \quad (n\ge 0)\,, \label{def_vn} \label{def_un}
\end{array}
\eea
$w_i(x,y)$ and $w_o(x,y)$ are the inhomogeneous solutions of
(\ref{scaledEqtau_2}) that vanish at $x=1$, and $p_n(x)$ and $q_n(x)$ will be derived below in
terms of the modified Bessel functions $I_0(x)$ and $K_0(x)$, and are normalized such that $p_n(1)=q_n(1)=1$.

The functions $v_n(y)$ and $u_n(y)$ satisfy the orthogonality relations
\bea
\begin{array}{c}
\ds \int_0^\beta u_n(y) u_m(y) dy = \int_0^\beta v_n(y) v_m(y)dy  = \frac{\beta}{2}\delta_{nm} \\
\ds \int_0^\beta v_n(y) u_m(y)dy = \frac{\beta}{2}\xi_{nm}\,,\label{defxinm}
\end{array}
\eea
where
\bea
\xi_{nm} =
\left\{
\begin{array}{l}
\ds \frac{2}{\beta}  \frac{l_n}{l_n^2 -k_m^2} = \frac{2}{\pi } \frac{(n+\frac{1}{2})}{(n+\frac{1}{2})^2 -m^2}
\,, \quad m\ge 1 \\ \\
\ds \frac{\sqrt 2}{\beta l_n} = \frac{\sqrt 2}{\pi } \frac{1}{n+\frac{1}{2} }\,,\quad m =0  \label{exprxinm}
\end{array}\right.
\eea
is an orthogonal matrix satisfying
\bea
\sum_{p=0}^\infty \xi_{pn} \xi_{pm} =  \sum_{p=0}^\infty \xi_{np} \xi_{mp} = \delta_{nm}\,. \nn
\eea
Using the orthogonality relations, we obtain the expansions
\bea\label{expansionun}
u_n(y)=\sum_{m=0}^\infty \xi_{mn} v_m(y) \,, \quad v_n(y)=\sum_{m=0}^\infty
\xi_{nm} u_m(y)\,.
\eea
%
\subsubsection{Derivation of $p_n(x)$ and $w_o(x,y)$}
%
The equation for $\hat \tau_o(x,y)$ in $\Omega_o$ is
\bea
\begin{array}{c}
\ds \( \frac{1}{x} \frac{\p}{\p x} x\frac{\p}{\p x} +
\frac{\p^2}{\p y^2}\) \hat \tau_o (x,y) = -\frac{1}{|\Omega|}\,, \quad (x,y) \in \Omega_o
\label{scaledEqtau_o} \\
\ds \frac{\partial}{\partial y}\hat \tau_o (x,y)\Big |_{y=0,\beta}=0\,, \quad
\frac{\partial}{\partial x}\hat \tau_o (x,y)\Big |_{x=\alpha} =0
\end{array}
\eea
and we choose $w_o(x,y)$ to satisfy
\bea
\begin{array}{c}
\ds  \( \frac{1}{x} \frac{\p}{\p x} x\frac{\p}{\p x} +
\frac{\p^2}{\p y^2}\) w_o(x,y) = -\frac{1}{|\Omega|} \,, \quad (x,y) \in \Omega_o \label{eqforv2} \nn\\
\ds w_o(1,y) = 0 \,, \quad \frac{\p }{\p y} w_o (x,y)\Big|_{y=0,\beta} =0 \,, \quad
\frac{\partial}{\partial x} w_o (x,y)\Big|_{x=\alpha} =0.
\end{array}
\eea
The solution for $w_o(x,y)$ is
\bea\label{wo}
w_o(x,y)=\frac{\ln x}{2\pi \beta} -\frac{x^2 -1}{4|\Omega|}\,.
\eea
Inserting $\hat \tau_o(x,y)$ from (\ref{deftauo}) into
(\ref{scaledEqtau_o}) yields for $p_n(x)$ the equations
\bea
\begin{array}{c}
\ds \( \frac{1}{x} \frac{\p}{\p x} x\frac{\p}{\p x} -k_n^2\) p_n(x) =0 \,,\quad  1 < x < \alpha \label{Equ2} \nn \\
\ds p_n(1) = 1\,, \quad \frac{\partial}{\partial x} p_n(x)\Big |_{x=\alpha} =0 \,.
\end{array}
\eea
Using the modified Bessel functions $I_0(x)$ and $K_0(x)$ and the
relations \cite{BookCarslawJaeger} ($I_0'(x)=I_1(x)$,
$K_0'(x)=-K_1(x)$,  we obtain
\bea
p_n(x) =  \frac{F_0(k_n x,k_n \alpha)}{F_0(k_n,k_n \alpha)}\,,
\label{uon}
\eea
with
\bea
F_0(x,y)=I_0(x)K_1(y)+K_0(x) I_1(y)\,.\nn
\eea
%
\subsubsection{Derivation of $q_n(x)$ and $w_i(x,y)$}
%
Proceeding similarly to the previous paragraph, the equation for
$\hat
\tau_i(x,y)$ in $\Omega_i$ is
\bea
\begin{array}{c}
\ds \( \frac{1}{x} \frac{\p}{\p x} x\frac{\p}{\p x} +
\frac{\p^2}{\p y^2}\) \hat \tau_i (x,y) = -\frac{1}{|\Omega|}\,, \quad (x,y) \in \Omega_i
\label{scaledEqtau_i} \\
\ds \hat \tau_i(x,0)= 0 \,, \quad
\frac{\partial}{\partial y}\hat \tau_i (x,y)\Big |_{y=\beta}=0\,.
\end{array}
\eea
We choose $w_i(x,y)$ to satisfy
\bea
\begin{array}{c}
\ds \( \frac{1}{x} \frac{\p}{\p x} x\frac{\p}{\p x} +
\frac{\p^2}{\p y^2}\) w_i(x,y) = -\frac{1}{|\Omega|}\,, \quad (x,y) \in \Omega_i \nn \\
\ds w_i(1,y) = 0\,,\quad w_i(x,0) = 0 \,, \quad
\frac{\partial}{\partial y}w_i(x,y)\Big |_{y=\beta} = 0 \,,
\end{array}
\eea
with solution
\bea
\begin{array}{lll}
\ds w_i(x,y) &=& \ds\frac{1}{|\Omega|} \sum_{n=1}^{\infty} c_n J_0(z_n x) \frac{\cosh(z_n
(\beta-y))}{\cosh(z_n
\beta)} -\frac{x^2 -1}{4|\Omega| } \\
\ds &=& \ds \frac{1}{|\Omega|} \sum_{n=1}^{\infty} c_n J_0(z_n x)\(  \frac{\cosh(z_n (\beta-y))}{\cosh(z_n
\beta)} - 1\)\,, \label{funcwi(x,y)}
\end{array}
\eea
where $z_n$ are the positive zeros of the Bessel function $J_0(x)$, and the coefficients $c_n$ are
given by
\bea\label{eqforcn}
c_n =\frac{2}{J_0'(z_n)^2} \int_0^1 J_0(z_n x)  \frac{x^2 -1}{4} x dx\,.
\eea
To derive expression (\ref{eqforcn}), we used the orthogonality
relation \cite{BookCarslawJaeger}
\bea
\int_{0}^{1} J_0(z_n x) J_0(z_m x)  x dx =  \delta_{nm} \frac{1}{2}\(
J_0(z_n)^2 + J_0'(z_n)^2 \). \nn
\eea
Inserting $\hat \tau_i(x,y)$ from (\ref{deftaui})
into (\ref{scaledEqtau_i}) gives for $q_n(x)$ the equation
\bea
\begin{array}{c}
\ds \( \frac{1}{x} \frac{\p}{\p x} x\frac{\p}{\p x} -l_n^2\) q_{n}(x) = 0\,,
\quad x < 1 \label{u1n} \nn \\
\ds q_{n}(1) = 1 \,,
\end{array}
\eea
and the solution that is regular at $x=0$ is
\bea
q_{n}(x) = \frac{I_0(l_n x)}{I_0(l_n)}\,.
\eea
%
\subsubsection{General expression for  $\hat \tau(x,y)$}
%
Using the expressions for $p_n(x)$, $q_n(x)$,
$w_i(x,y)$ and $w_o(x,y)$, the NET solution is
\bea\label{tauexpansion1}
\hat \tau(x,y)=\left\{
\begin{array}{l}
\displaystyle  \hat \tau_i(x,y) = \sum_{n=0}^\infty  b_n \frac{I_0(l_n x)}{I_0(l_n)} v_n(y) + w_i(x,y) \,, \quad x\le 1 \\ \\
\displaystyle \hat \tau_o(x,y)  = \sum_{n=0}^\infty  a_n \frac{F_0(k_n x,k_n \alpha)}{F_0(k_n,k_n \alpha)} u_n(y) +
\frac{\ln x}{2\pi \beta} -\frac{x^2 -1}{4|\Omega|}
\,, \quad 1 \le x \le  \alpha,
\end{array}\right.
\eea
where the unknown coefficients $a_n$ and $b_n$ will be determined
by patching the two expressions at $x=1$. The continuity condition
for $\hat \tau(x,y)$ at $x=1$ gives
\bea\label{matchcondtau}
\sum_{n=0}^\infty a_n   u_n(y) =  \sum_{n=0}^\infty b_n  v_n(y)\nn\,,
\eea
and using the expansions in (\ref{expansionun}), we obtain that
$a_n$ and $b_n$ are related by
\bea\label{an_bn}
b_n =\sum_{m=0}^\infty \xi_{nm} a_m  \,, \quad a_m= \sum_{n=0}^\infty \xi_{nm}  b_n\,.
\eea
The continuity condition for the flux at $x=1$ gives
\bea
\sum_{n=0}^\infty  b_n l_n \frac{I_1(l_n)}{I_0(l_n)} v_n(y)  -
 \sum_{n=1}^\infty  a_n k_n \frac{F_1(k_n,k_n \alpha)}{F_0(k_n,k_n \alpha)} u_n(y)
= \frac{1}{2\pi \beta}- \frac{1}{2|\Omega|}- \frac{\p}{\p x}w_i(x,y)\Big |_{x=1} \label{condfromderivative1}\,, \nn
\eea
with
\bea
F_1(x,y)=\frac{\p}{\p x} F_0(x,y) = I_1(x)K_1(y) - K_1(x) I_1(y)\,. \nn
\eea
 This can be  rewritten as
\bea
\sum_{n=0}^\infty  b_n \beta_n v_n(y) + \sum_{n=0}^\infty  a_n \alpha_n u_n(y)  =
\sum_{n=0}^\infty  \gamma_n  u_n(y) \,. \label{condfromderivative3_1}
\eea
where
\bea
\begin{array}{c}
\ds \alpha_0 = 0 \,, \quad \alpha_n= -k_n \frac{F_1(k_n,k_n \alpha)}{F_0(k_n,k_n \alpha)}\,\, ( n \ge 1) \,, \quad  \label{defalphan}
\ds \beta_n = l_n \frac{I_1(l_n)}{I_0(l_n)} \label{defbetan}
\end{array}
\eea
and the $\gamma_n$ are implicitly defined by the equation
\bea
\sum_{n=0}^\infty  \gamma_n u_n(y)
= \frac{1}{2\pi \beta}- \frac{1}{2|\Omega|}- \frac{\p}{\p x}w_i(x,y)\Big |_{x=1}\,.
\label{defgamman}
\eea
By using the expansions in (\ref{expansionun}) we obtain from (\ref{condfromderivative3_1})
\bea
\begin{array}{lll}
\ds \sum_{m=0}^\infty \alpha_m  \xi_{nm} a_m  + \beta_n b_n &=&  \ds \sum_{m=0}^\infty  \xi_{nm}  \gamma_m  \label{eqforanban1} \\
\ds \alpha_n  a_n  + \sum_{m=0}^\infty \beta_m b_m \xi_{mn} &=& \ds \gamma_n \,. \label{eqforanban2}
\end{array}
\eea
Finally, using the relations between $a_n$ and $b_n$ given in (\ref{an_bn}), we obtain the matrix equations
\bea\label{eqforanmatrix}
\begin{array}{lll}
\ds \sum_{m=0}^\infty (\beta_n + \alpha_m ) \xi_{nm} a_m &=&   \ds \sum_{m=0}^\infty \xi_{nm} \gamma_m \\
\ds \sum_{m=0}^\infty (\beta_m + \alpha_n) \xi_{mn} b_m &=& \ds \gamma_n  \,.
\end{array}
\eea
For given $\alpha$ and $\beta$, by truncating and numerically solving these equations we find approximated values for $a_n$ and $b_n$, and from this we obtain an approximation for $\hat \tau(x,y)$. We will analyze the equations for $a_n$ and $b_n$ in more detail later on.
%
\subsection{MFPT with a uniform initial distribution}
%
We shall first consider the average MFPT $\hat \tau(x)$ when the Brownian particle is
initially uniformly distributed at radial position $x$. Using (\ref{tauexpansion1}) we obtain
\bea\label{meanexittime1}
\hat \tau(x) =  \frac{1}{\beta}\int_0^\beta \hat \tau(x,y)dy = \left\{
_{}\begin{array}{l}
\displaystyle    \hat \tau_i(x)  = \frac{1}{\beta} \sum_{n=0}^\infty  \frac{b_n}{l_n} \frac{I_0(l_n x)}{I_0(l_n)} +
 \frac{1}{\beta} \int_0^\beta w_i(x,y) dy \,,\quad    x \in \Omega_i \\ \\
\displaystyle    \hat \tau_o(x)   = \frac{a_0}{\sqrt 2}  +
\frac{\ln x}{2\pi \beta } - \frac{x^2-1}{4 |\Omega|}
\,, \quad x \in \Omega_o.
\end{array}\right.
\eea
Expression (\ref{meanexittime1}) shows that $\frac{a_0}{\sqrt 2}$
is the averaged MFPT for Brownian particles that are initially
uniformly distributed at $x=1$. The expression for $\hat
\tau_o(x)$ has an intuitive interpretation: the escape time starting
at $x\ge 1$ is the sum of the average time to reach $x=1$, plus the
escape time starting at $x=1$.

The average MFPT $\hat \tau$ for particles that are
initially uniformly distributed in $\Omega$ is
\bea
\hat \tau &=& \frac{1 }{|\Omega|} \int_{\Omega} \hat \tau(x,y) dV  = \frac{|\Omega_i|}{|\Omega|} \frac{1 }{|\Omega_i|}\int_{\Omega_i} \hat \tau(x,y) dV +\frac{|\Omega_o|}{|\Omega|} \frac{1 }{|\Omega_o|}\int_{\Omega_o} \hat \tau(x,y) dV \nn\\
&=& \frac{|\Omega_i|}{|\Omega|} \hat \tau_i + \frac{|\Omega_o|}{|\Omega|} \hat \tau_o = \frac{1}{\alpha^2 }\hat \tau_i +  \frac{\alpha^2-1}{\alpha^2} \hat \tau_o\,,
\label{averagehattau}
\eea
where
\bea \label{meanexittime2}
\begin{array}{lll}
\ds \hat \tau_i  &=& \ds  \frac{2}{\beta} \sum_{n=0}^\infty  \frac{b_n}{l_n^2}
\frac{I_1(l_n)}{I_0(l_n)} +  \frac{2}{\beta} \int_0^1 \int_0^\beta w_i(x,y) x dx dy \\
\ds \hat \tau_o  &=& \ds{\frac{a_0}{\sqrt 2} + \frac{\alpha^2}{\alpha^2-1 }\frac{\ds{4\ln \alpha -3 + \frac{4}{\alpha^{2}}-\frac{1}{\alpha^{4}}}}{8\pi \beta}}\,.
\end{array}
\eea
The time $\hat \tau_i$ is the average MFPT for particles starting uniformly distributed in the inner cylinder $\Omega_i$, and $\hat \tau_o$ is the average MFPT for particles starting uniformly distributed in the annulus $\Omega_o$. We shall now derive asymptotic limits for $\hat \tau_i$ and $\hat \tau_o$ under various conditions.

%
\subsection{Asymptotic expressions for a cylinder with $R \gg a$, and a flat cylinder
with $R\gg a$ and $h \ll R$}
%
We will first derive asymptotic expressions for $\hat \tau$ for a cylinder with $R \gg a$ ($\alpha \gg 1$) and arbitrary height $h$, and we will then focus on a flat cylinder with $h\ll R$ ($\beta \ll \alpha$). We show that $a_0(\alpha,\beta)$ is the leading order contribution to
$\hat \tau$ for $\alpha \gg 1$. For a flat cylinder with $\alpha
\gg 1$ and $\beta/\alpha \ll 1$, we further have that $a_0(\alpha,\beta)
\approx a_0(\beta)$. To derive $a_0(\beta)$ as a function of
$\beta$, we consider the limit $\alpha \to \infty$ while
$\beta$ stays bounded ($R\to \infty$ with finite $h$). We show that $\hat \tau(x,y)$ and $\hat \tau$ have finite asymptotic limits for $\alpha \to \infty$ that depend only on $\beta$.

We start by considering the limit $\alpha \gg 1$. The function
$w_i(x,y)$ in (\ref{funcwi(x,y)}) is of the order
$|\Omega|^{-1}\sim \alpha^{-2}$ and can be neglected, and we have
\beq
\hat \tau_i(x,y) \approx \sum_{n=0}^\infty  b_n \frac{I_0(l_n
x)}{I_0(l_n)} v_n(y).
\eeq
Because the average time $\hat \tau_i$ starting uniformly
distributed in $\Omega_i$ is similar to the the average time $\hat
\tau(1)=\frac{a_0}{\sqrt 2}$ starting uniformly distributed at
$x=1$, the contribution of $\tau_i$ in (\ref{averagehattau}) is by
a factor $\alpha^{-2}$ smaller compared to the contribution of $\hat
\tau_o$, and we arrive at the asymptotic expression
\bea
\hat \tau \approx \hat \tau_o \approx \frac{a_0(\alpha,\beta)}{\sqrt 2}  + \frac{4\ln \alpha -3 }{8\pi \beta} \,. \label{meanexittime3}
\eea
The dimensional time $\tau$ is
\bea\label{meanexittime3_1}
\tau \approx \frac{|V|}{aD} \hat \tau \approx \frac{|V|}{aD}\frac{a_0(\alpha,\beta)}{\sqrt 2}  + \frac{R^2}{8D} \(4\ln\(\frac{R}{a}\) -3 \)\,.
\eea
In particular, for $\beta\gg \ln\alpha$ and $\alpha \gg 1$ we obtain the result
\bea\label{meanexittime4}
\hat \tau \approx \frac{a_0(\alpha,\beta)}{\sqrt 2}  \quad \Longrightarrow \quad \tau \approx \frac{|V|}{aD}\frac{a_0(\alpha,\beta)}{\sqrt 2}\,.
\eea
Equations (\ref{meanexittime3_1})-(\ref{meanexittime4}) show that
$a_0(\alpha,\beta)$ is the leading term that determines the average MFPT for
$\alpha\gg1$. To further evaluate $\tau$, we shall now estimate $a_0(\alpha,\beta)$ for a flat cylinder with a small hole, when $\beta \ll \alpha$ and $\alpha \gg 1$, by considering the
limit $\alpha \to \infty$ while $\beta$ remains finite ($R\to \infty$ with fixed $h$). For $\alpha \to \infty$, the scaled times $\hat \tau(x,y)$ and $\hat \tau$ have have finite limits that depend on $\beta$, and only the dimensional times $\tau(r,z)$ and $\tau$ diverge $\sim R^2$. In this limit, the
coefficients $\alpha_n$, $\beta_n$ and $\gamma_n$ in
(\ref{defalphan}) and (\ref{defgamman}) are given by
\bea\label{asympalphanbetantaulargealpha}
\alpha_0=0, \quad \alpha_n = k_n \frac{K_1(k_n)}{K_0(k_n)}\,(n\ge 1)\,, \quad
\beta_n = l_n \frac{I_1(l_n)}{I_0(l_n)} \,, \quad
\gamma_n = \frac{\delta_{n0}}{\sqrt 2 \pi \beta} = \gamma_0 \delta_{n0} \,,
\eea
with $\gamma_0=\frac{1}{\sqrt 2 \pi \beta}$, and
(\ref{eqforanban1}) simplifies to
\bea\label{equa_n}
\sum_{m=0}^\infty (\beta_n + \alpha_m ) \xi_{nm} a_m =
 \xi_{n0} \gamma_0 \label{eqanasymp} \,.
\eea
$\alpha_n$, $\beta_n$ and $\gamma_n$ are functions of $\beta$
only, and hence, also $a_n$ and $b_n$ depend only on
$\beta$. For $\alpha \gg 1$ and $\beta /\alpha \ll 1$, $\hat \tau(x,y)$ in (\ref{tauexpansion1}) is in
first order given by
\bea\label{tauinlimitalpahtoinfty}
\hat \tau(x,y)=\left\{
\begin{array}{l}
\displaystyle  \sum_{n=0}^\infty  b_n(\beta) \frac{I_0(l_n x)}{I_0(l_n)} v_n(y) \,, \quad x\le 1 \\ \\
\displaystyle \frac{a_0(\beta)}{\sqrt 2} + \sum_{n=1}^\infty  a_n(\beta) \frac{K_0(k_n x)}{K_0(k_n)} u_n(y)
+ \frac{\ln x}{2\pi \beta } \,, \quad  1 \le x \ll \alpha  \,,
\end{array}\right.
\eea
where we used
\bea
\frac{F_0(k_n x,k_n \alpha)}{F_0(k_n,k_n \alpha)} \approx \frac{K_0(k_n x)}{K_0(k_n)}\,, \quad \alpha \gg 1 \mbox{ and } x \ll \alpha \,. \nn
\eea
We conclude that the NET for $\alpha \gg 1$ and $\beta /\alpha \ll
1$ is in leading order
\bea
\tau \approx \frac{|V|}{aD}\frac{a_0(\beta)}{\sqrt 2}  + \frac{R^2}{8D} \(4\ln\(\frac{R}{a}\) -3 \)\,.
\label{meanexittime3_smallbeta}
\eea
In the next section we shall analyze the behavior of $a_0$ as a function of $\beta$.
%
\subsubsection{Behavior of ${a_0(\beta)}$ as a function of $\beta$}
\label{subsubsection_a0}
To evaluate $a_0(\beta)$ as a function of $\beta$, we solve numerically (\ref{eqanasymp}) by
truncating the series at sufficiently high values $n$: in Fig.~\ref{fig_a0}(a) we plot the analytic approach result for ${a_0(\beta)}/{\sqrt 2}$ and confirm that it agrees well with results from Brownian simulations that were performed with $\alpha=50$. Interestingly, Fig.~\ref{fig_a0}a shows that the simulation result for $\beta = 40$ (when $\beta$ is comparable to $\alpha=50$) still agrees very well with the analytic
result derived with the assumption $\alpha \gg \beta$, suggesting that $a_0(\beta)$
remains a good approximation until values $\beta \sim
\alpha$ ($h \sim R$). As a consequence, this suggests that
(\ref{meanexittime3_smallbeta}) is an acceptable approximation for $\tau$
until values $h \sim R$. Fig.~\ref{fig_a0}a shows that $a_0/\sqrt{2}$ approaches the value
$\frac{1}{4}$ for large $\beta$, thus, from (\ref{meanexittime4}) we recover the narrow escape formula $\tau \approx \frac{|V|}{4aD}$ \cite{Ward1,Grigorievetal2002,HolcmanetalNE1} derived for a volume
with isoperimetric ratio of order 1. Conversely, (\ref{meanexittime3})
shows that the validity of the narrow escape formula
$\frac{|V|}{4aD}$ is not limited to the range where
$h\sim R$, but it is already a valid approximation when $\ln \alpha \ll
\beta$ and $\beta \gtrsim 40$ (Fig.~\ref{fig_a0}a). Hence, we
conclude that $\tau = \frac{|V|}{4aD}$ is a good
approximation even for an oblate volume with $R \gg h$ (and $h\gg
a$). In the opposite limit $\beta \to 0$, we find that
${a_0(\beta)}/{\sqrt 2}$ does not converge towards zero
(Fig.~\ref{fig_a0}a), but $\lim_{\beta \to 0}{a_0(\beta)}/{\sqrt
2} \approx 0.071$ (in appendix \ref{asymptoticForBetaTo0} we derive an
analytical approximation for $a_0(\beta)$ for $\beta \to 0$). In Fig.~\ref{fig_a0}b we show the effect of the truncation level $n$ on the value of ${a_0(\beta)}/{\sqrt 2}$: after $n \sim 100$ the steady state regime is achieved.{ Finally, in Fig.~\ref{fig_a0}c-d, we compare the values of the coefficients $a_0(\beta)$ and $b_0(\beta)$, where $b_0(\beta)$ is obtained using (\ref{an_bn}). The graphs show that $b_0(\beta)\approx a_0(\beta)$ is a good approximation, and we will use this in section \ref{section_truncation}.
}

\begin{figure}[h!]
  \begin{center}
    \subfigure[]{\includegraphics[scale=0.35]{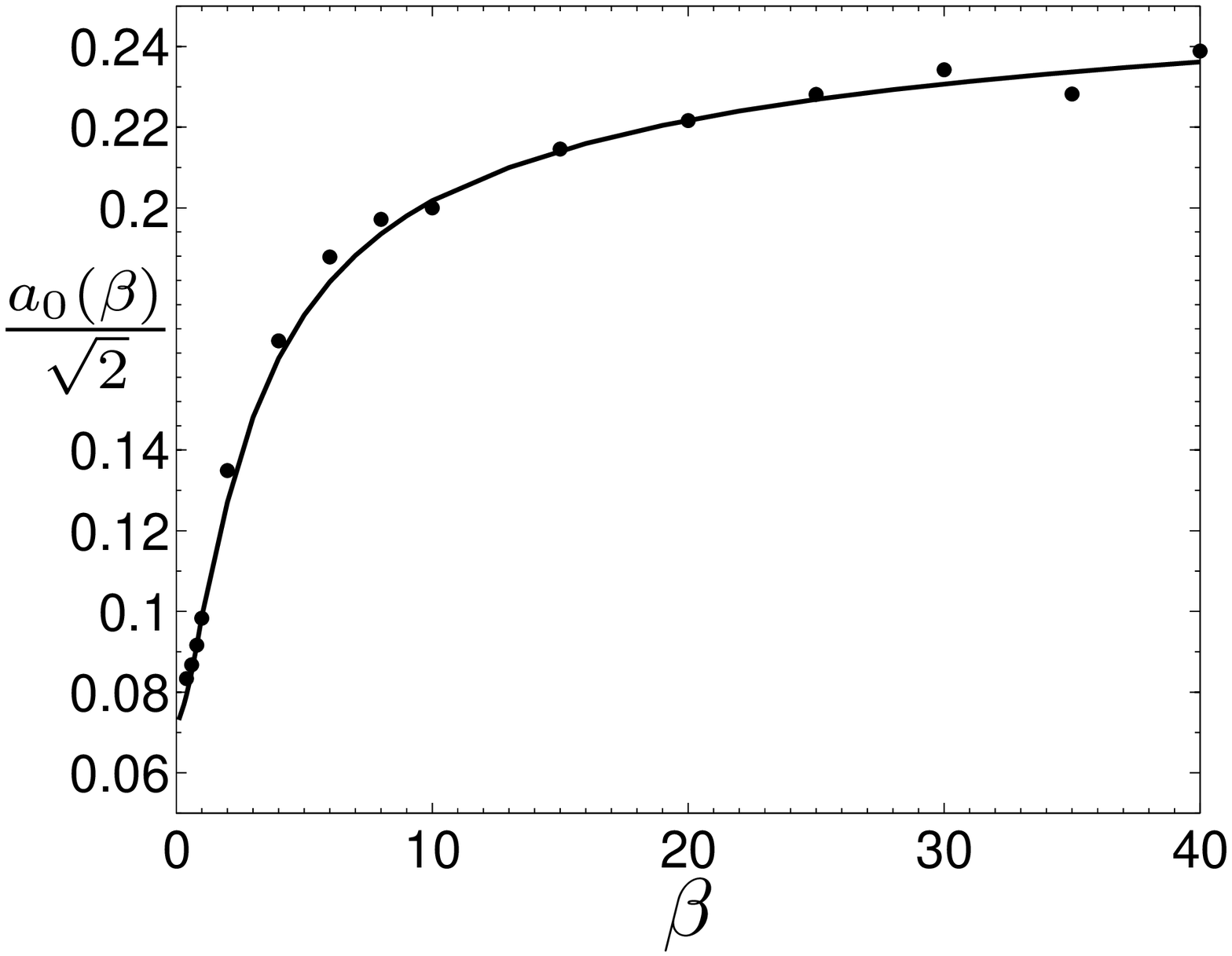}}
    \subfigure[]{\includegraphics[scale=0.49]{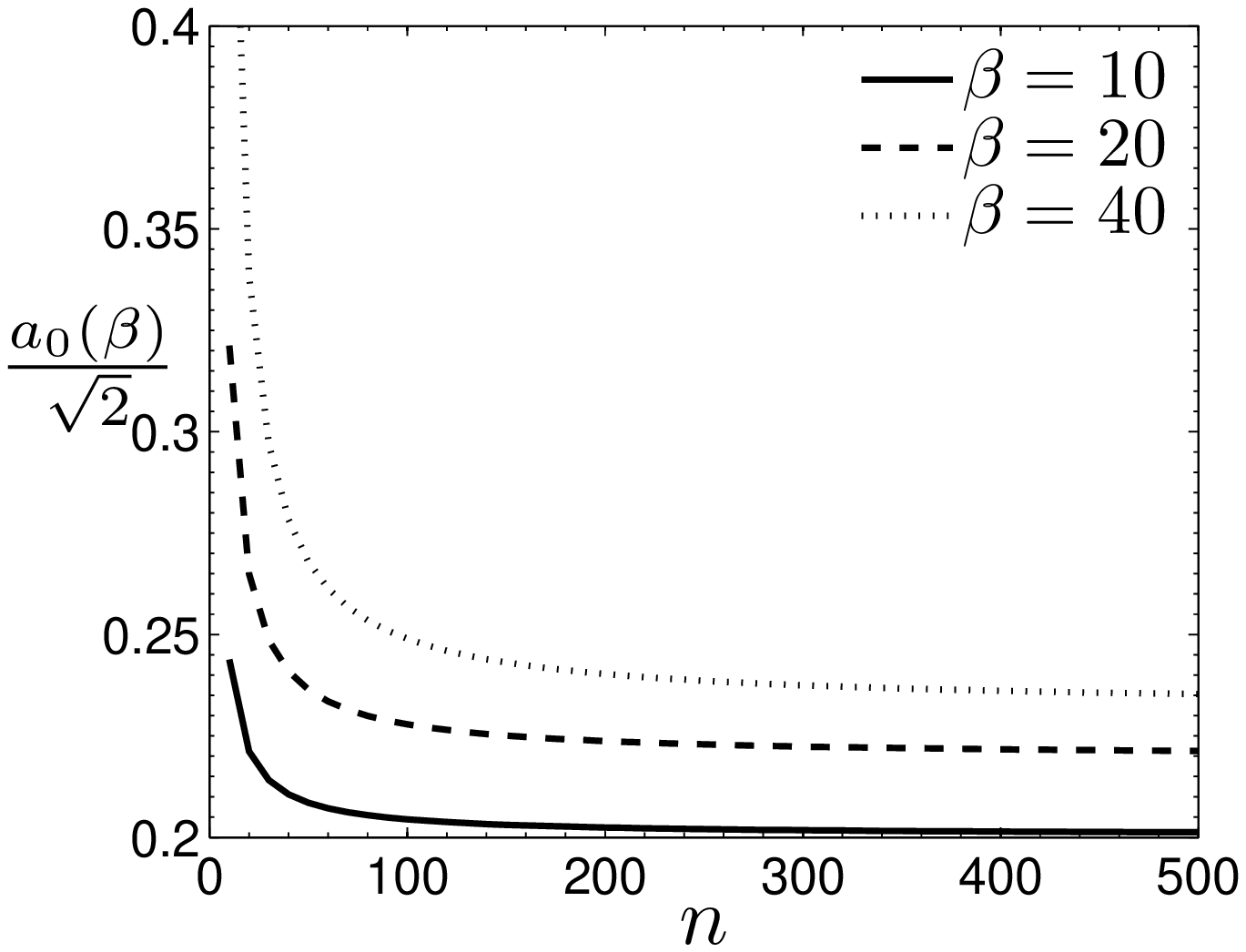}}
    \subfigure[]{\includegraphics[scale=0.49]{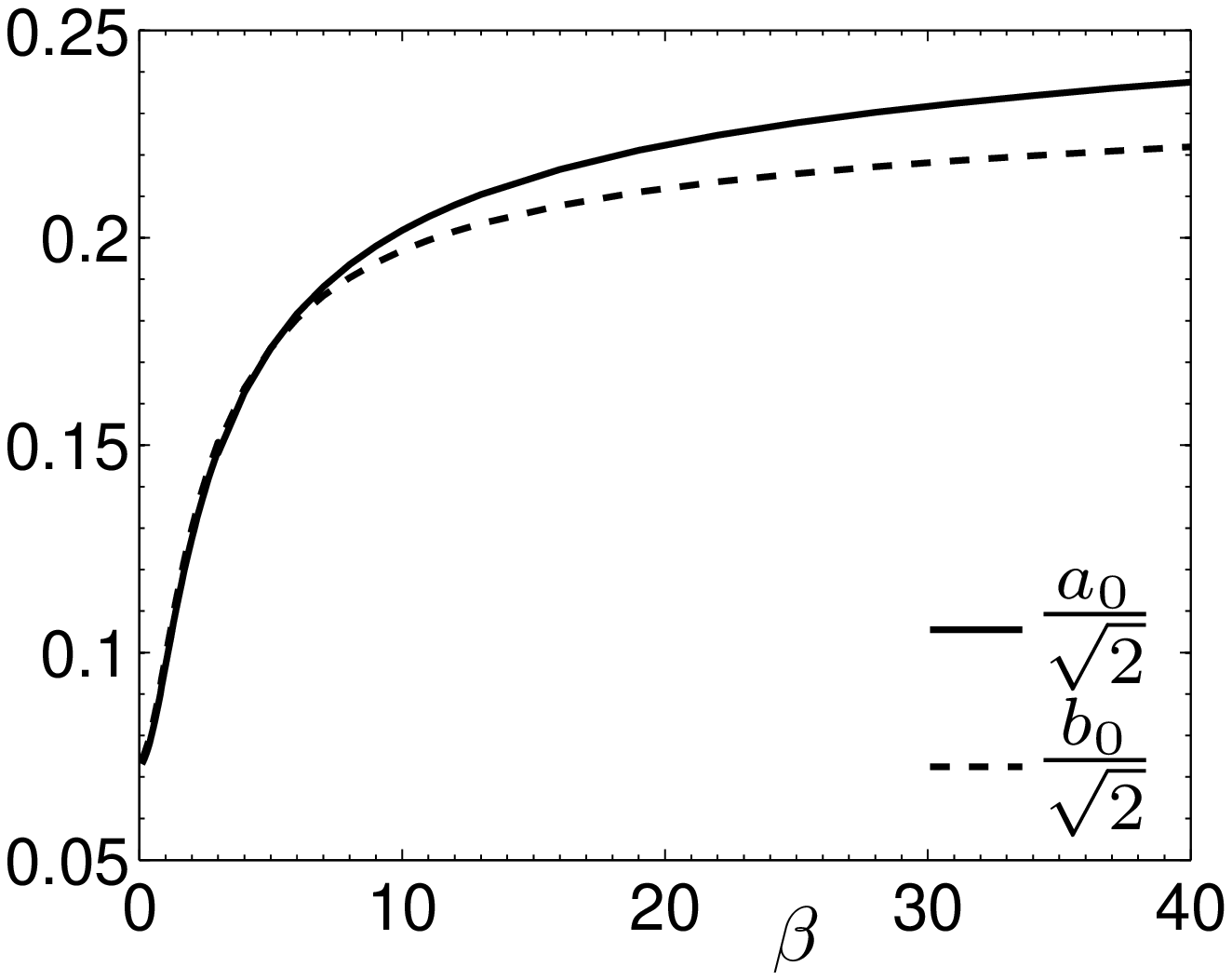}}
     \subfigure[]{\includegraphics[scale=0.49]{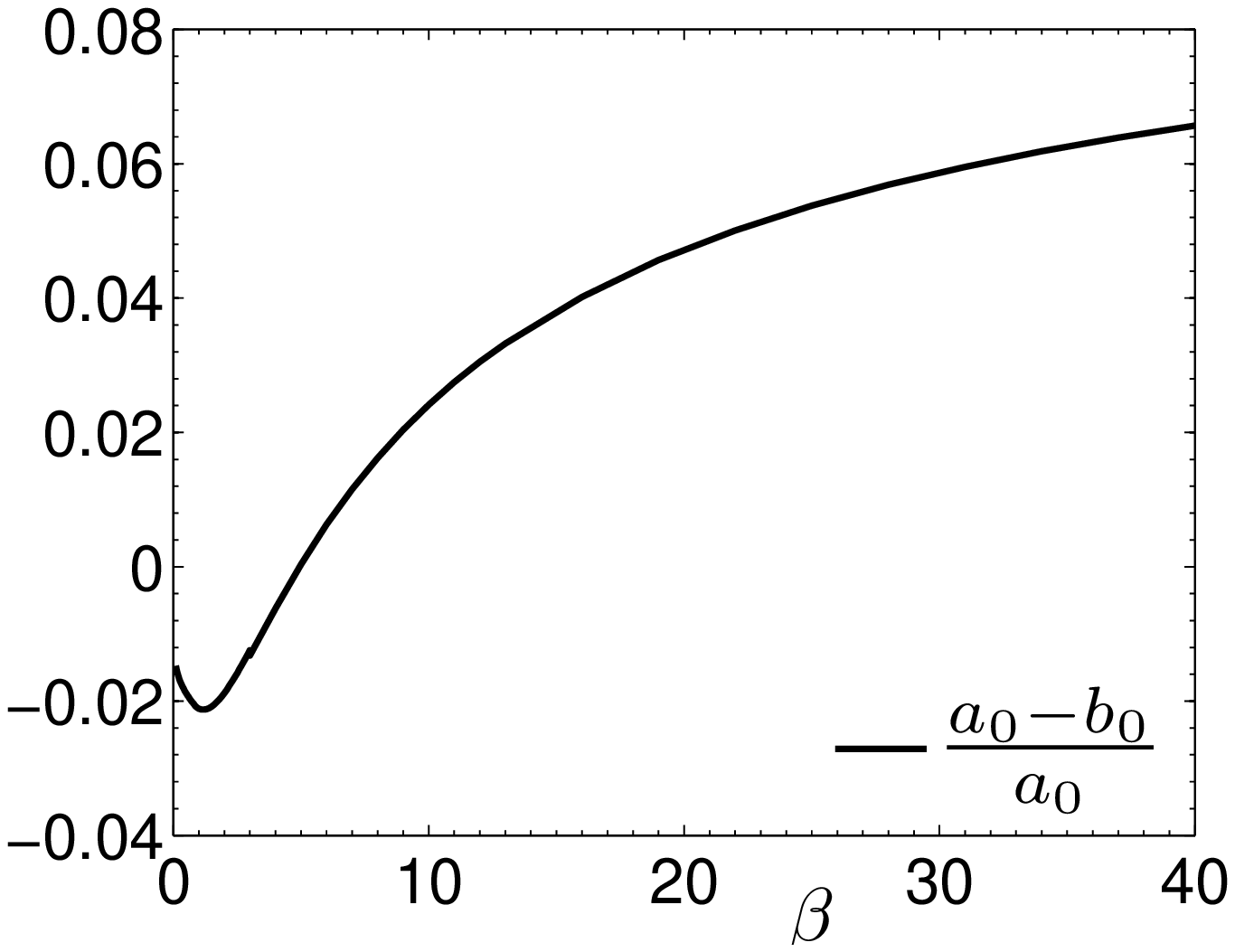}}
  \end{center}
  \caption{(a) Graph of $\frac{a_0(\beta)}{\sqrt{2}}$  as a function of $\beta$ for $\alpha \gg 1$.
  The values for $a_0$ are obtained by truncating and numerically solving (\ref{eqanasymp}) (for $\beta>10$ we truncate at $n=500$). ${a_0(\beta)}/{\sqrt {2}}$ is the average MFPT for Brownian particles starting uniformly distributed at $x=1$. The analytic estimations for ${a_0(\beta)}/{\sqrt {2}}$ are compared to results from Brownian simulations (data points) obtained for 10 000 Brownian trajectories, starting uniformly distributed at $x=1$ ( for $\alpha=50$). (b) Dependency of $\frac{a_0(\beta)}{\sqrt{2}}$ on the truncation level $n$ for various $\beta$. (c) Comparison between $a_0(\beta)$ and $b_0(\beta)$, where $b_0$ is obtained from (\ref{an_bn}), showing that $a_0(\beta)\approx b_0(\beta)$. (d) Relative difference $(a_0(\beta)-b_0(\beta))/a_0(\beta)$.
  }
\label{fig_a0}
\end{figure}

%
\subsubsection{Boundary layer analysis: particles starting near the absorbing hole}
%
In the neighborhood of the small absorbing window (for $x\sim 1$ and $y\sim1$), there is a
boundary layer (BL) where the behavior of $\hat \tau(x,y)$ is very different compared to large $x$ and $y$. In Fig.~\ref{fig_tau_boundlayer} we study numerically the
shape of the BL using (\ref{tauinlimitalpahtoinfty}). The different panels depict $\hat\tau(x,y)$ in the neighborhood of the absorbing window for various $\beta$. The plots show that a boundary layer starts to evolve around $\beta \sim 0.3$, and the evolution is almost finished for $\beta \sim 10$ (there is no significant difference between the plots for $\beta=10$ and $\beta=50$). Furthermore, the approximate extent of the boundary layer for large $\beta$ is $\Delta x \sim \Delta  y \sim 10$.

In Fig.~\ref{fig_MFPT}a, we show $\hat \tau(x,\beta)$ for particles released on the upper surface at $y=\beta$ as a function of $x$ and for various $\beta$. Such a situation is relevant at
synapses where neurotransmitters are released at the presynaptic
terminal, located opposite to the surface with the postsynaptic density (PSD)
where receptors are clustered \cite{Ehlers_Review_GutamateRec_2008,Harris_Review_Spine_2008}.
The hole radius $a$ would correspond to the radius of the PSD. Fig.~\ref{fig_MFPT}(a) shows that when the height of the synaptic cleft is comparable to the radius of the PSD ($\beta \sim
1)$, $\hat \tau(x,\beta)$ changes considerably as a function of the radial release position $x$.
In contrast, for $\beta\gg 1$ the release site is outside the boundary layer and the NET is almost
independent of $x$ and is well approximated by $1/4$.

\begin{figure}[h!]
\begin{center}
      \includegraphics[scale=0.30]{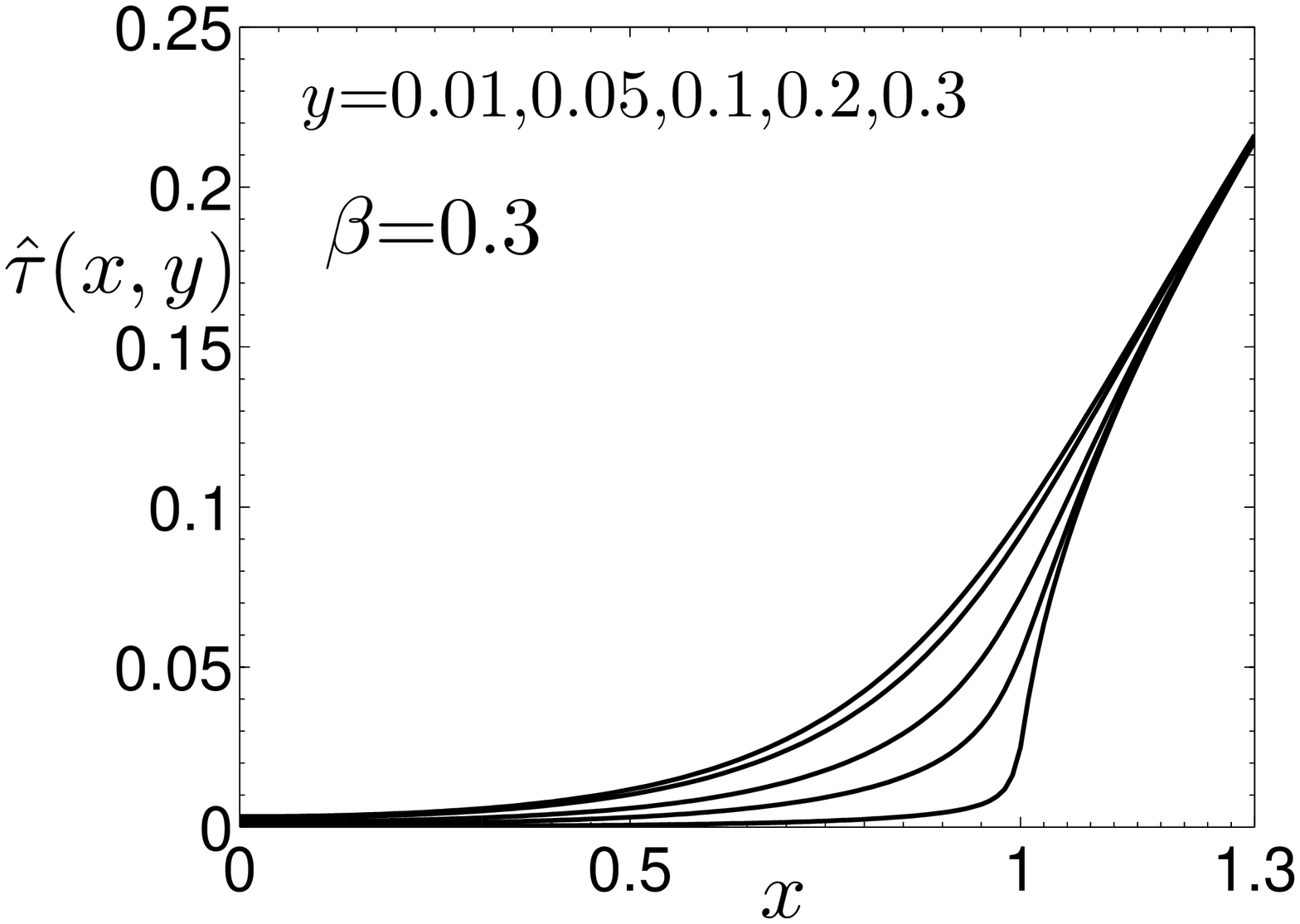} \hspace{0.5cm}
       \includegraphics[scale=0.30]{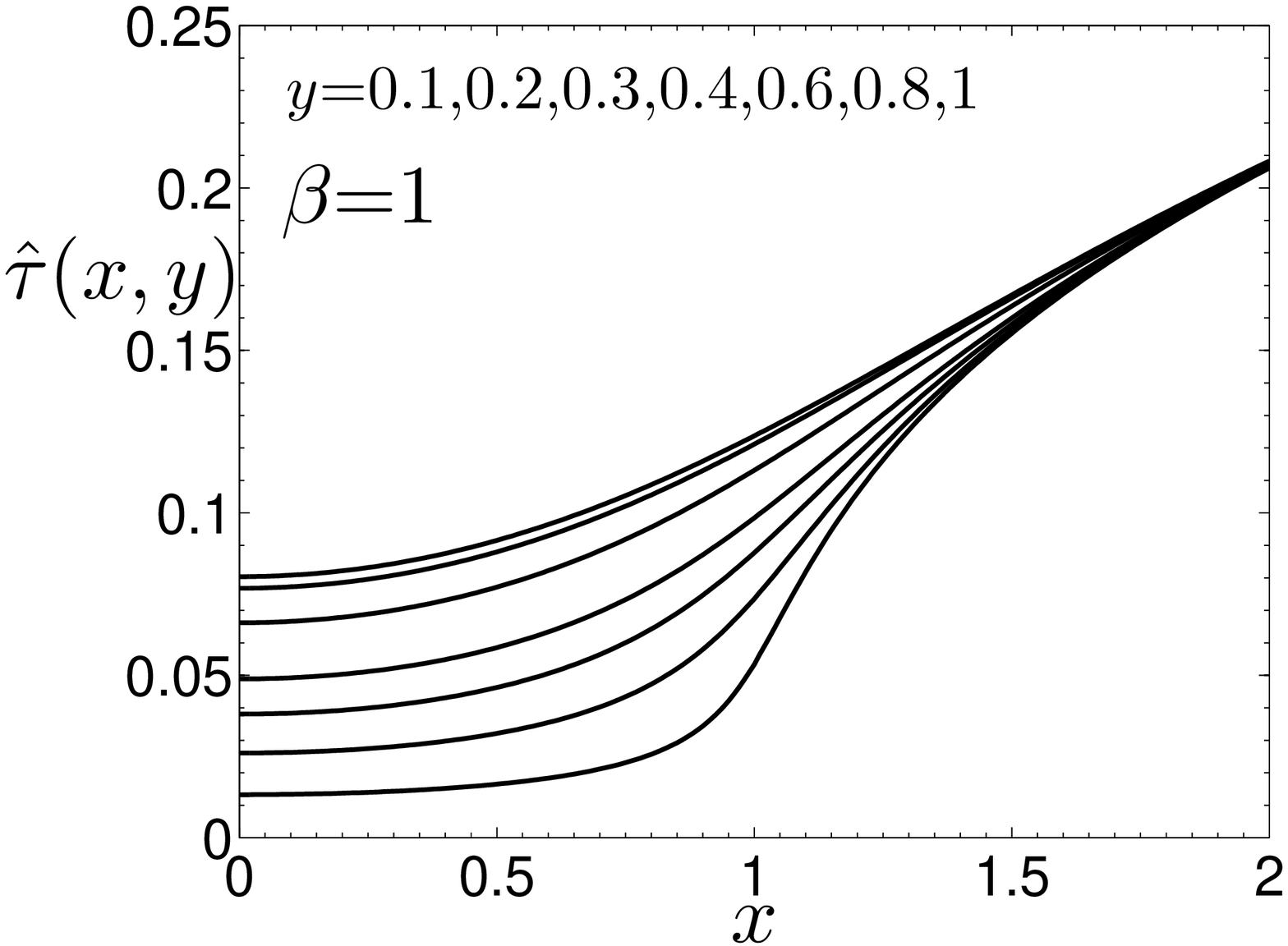} \\
      \includegraphics[scale=0.30]{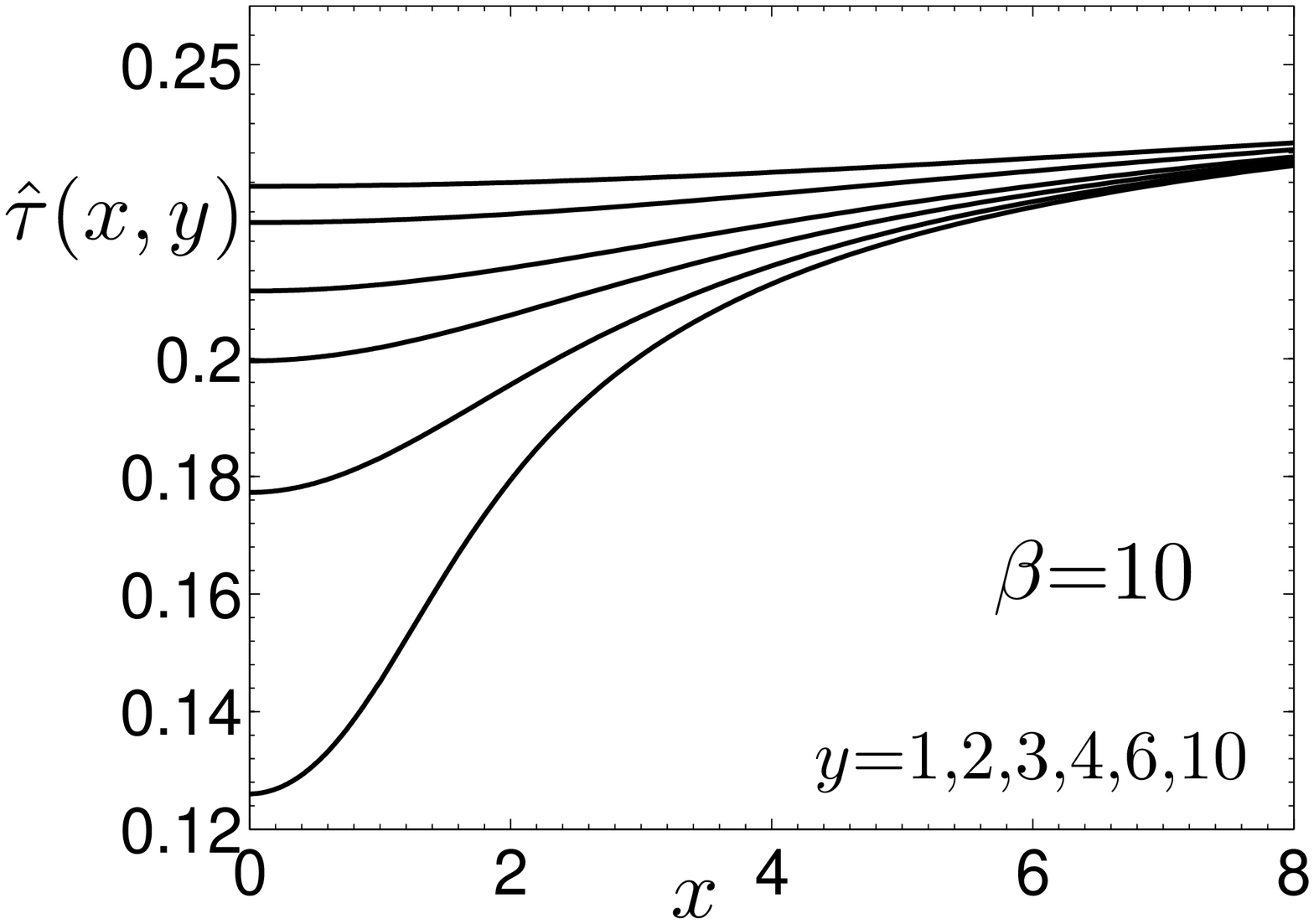} \hspace{0.5cm}
       \includegraphics[scale=0.305]{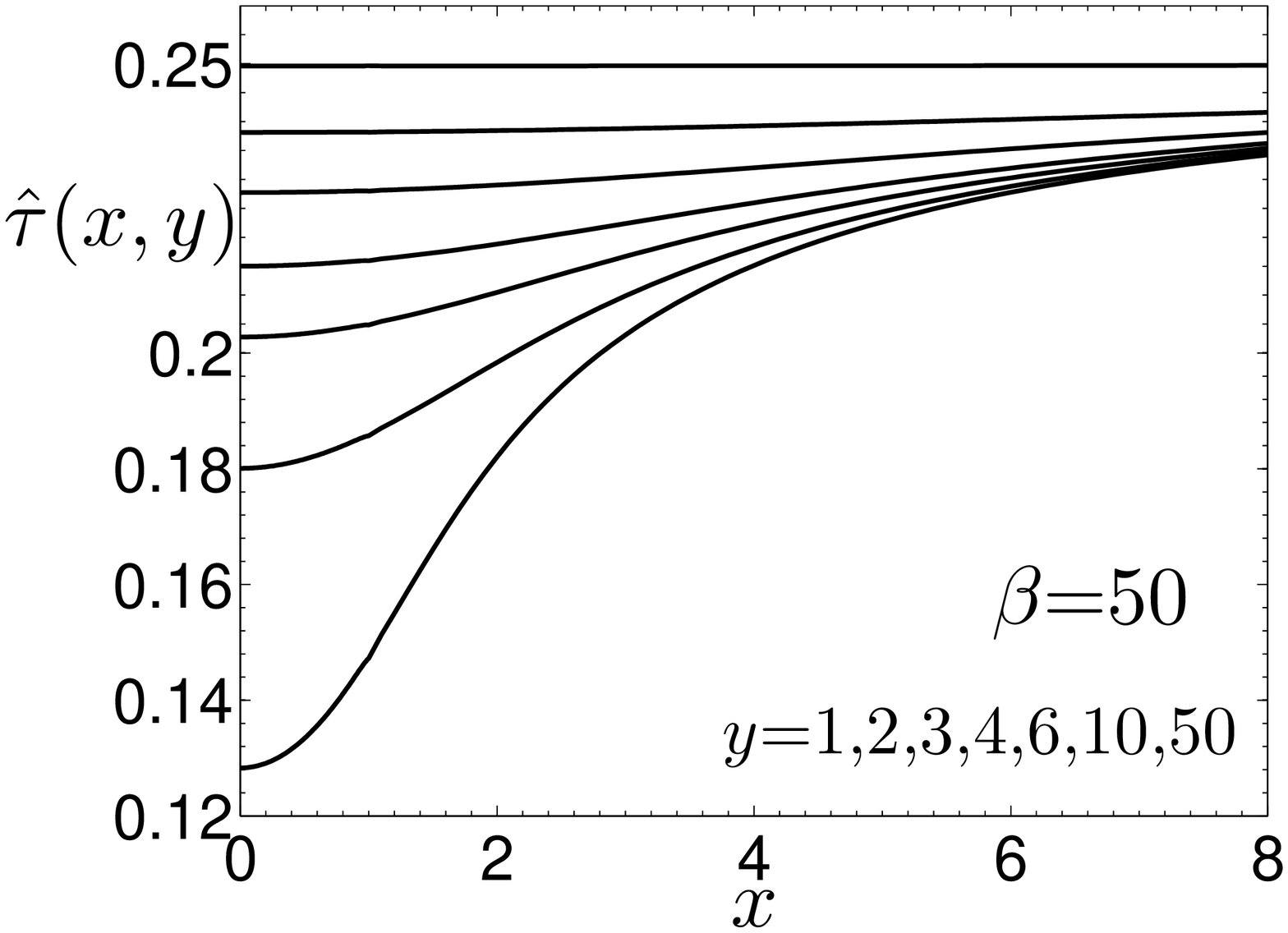}
       \caption{Shape of the boundary layer at the absorbing hole for $\alpha \gg 1$ and different values of $\beta$.
       The NET $\hat \tau(x,y)$ (from (\ref{tauinlimitalpahtoinfty})) as a function of $x$
       for different $y$ and $\beta$ (as displayed in each panel). }
  \label{fig_tau_boundlayer}
\end{center}
\end{figure}

\begin{figure}[h!]
\begin{center}
       \subfigure[]{\includegraphics[scale=0.36]{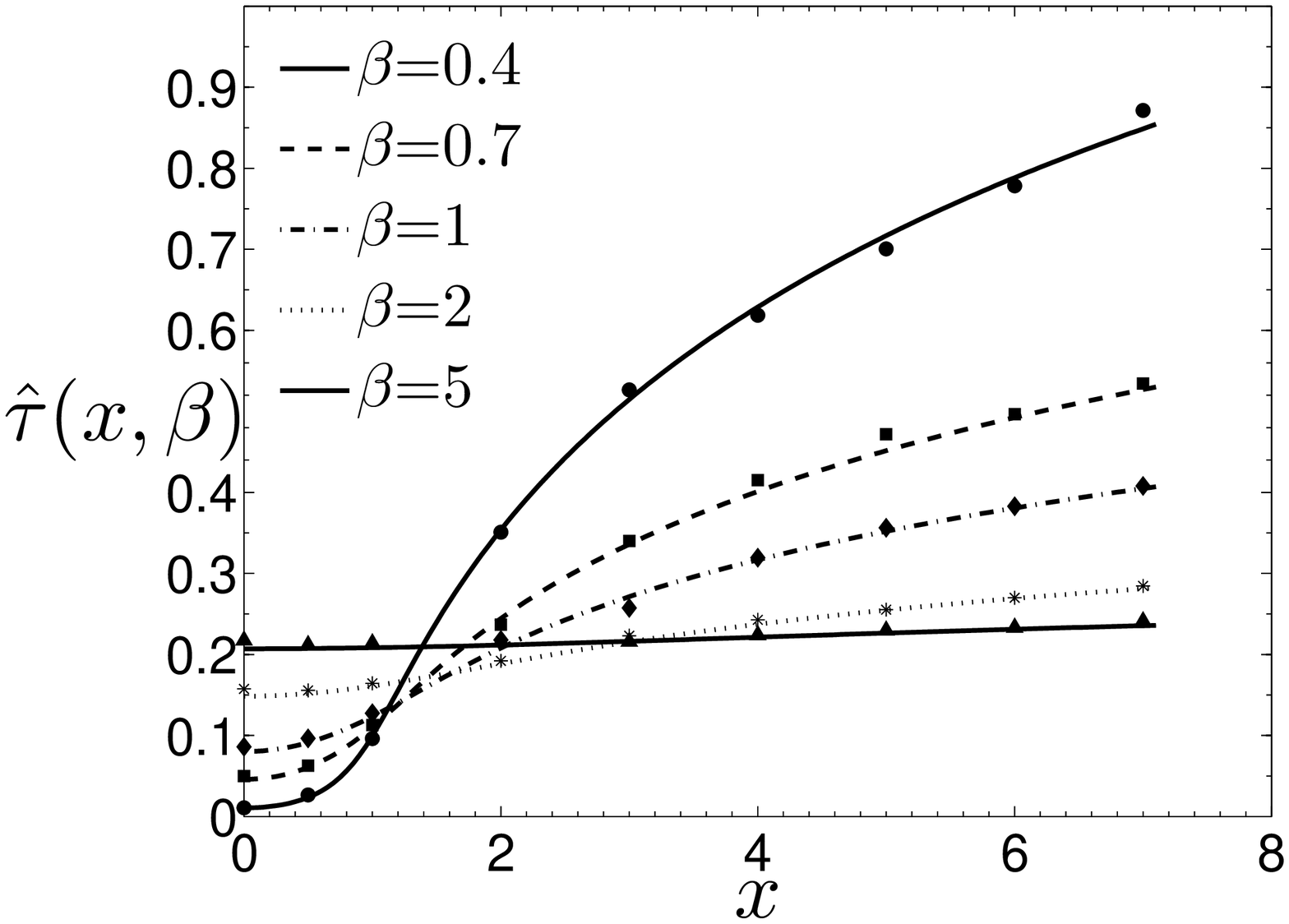}}
       \subfigure[]{\includegraphics[scale=0.48]{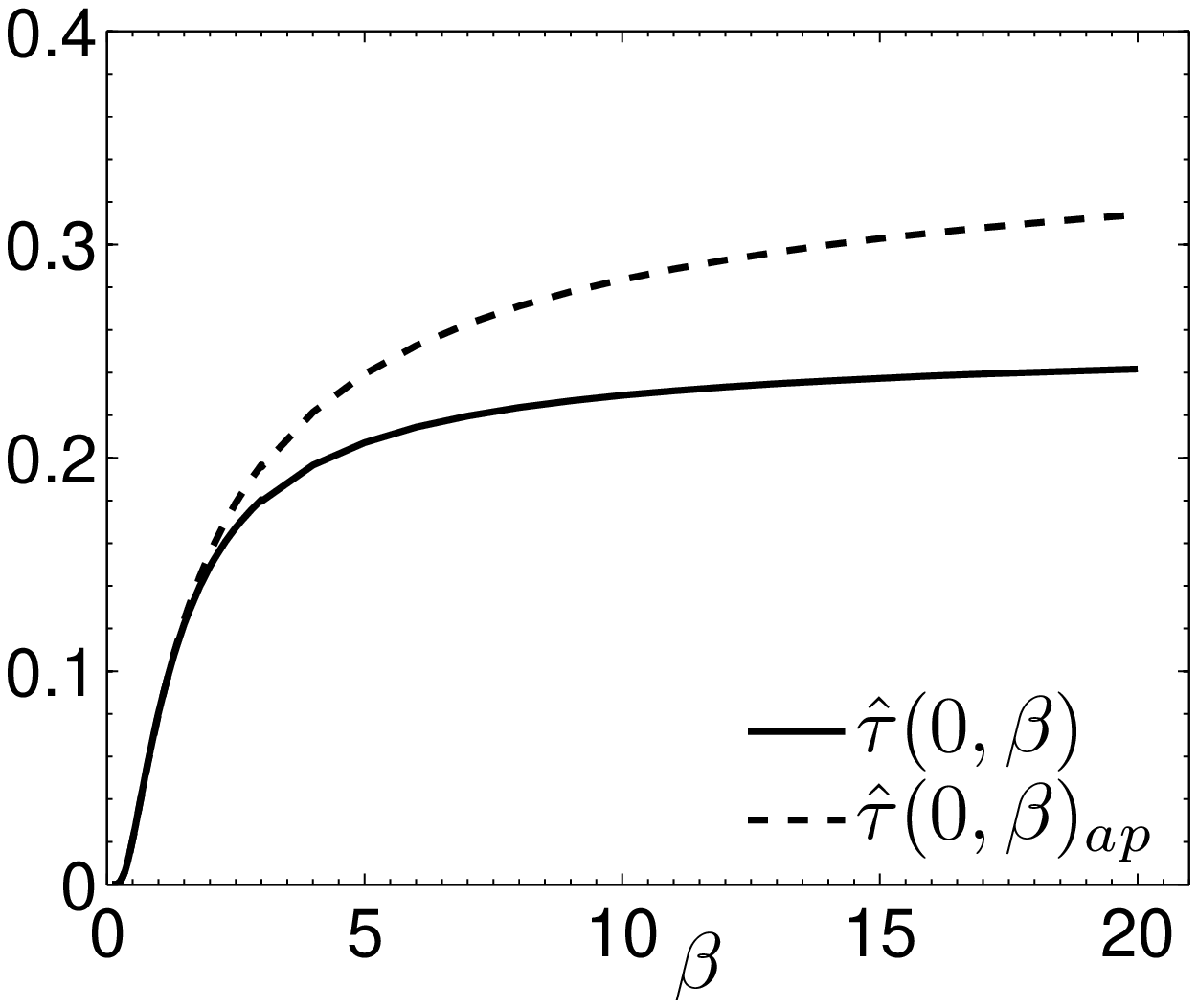}}
       \caption{(a) The NET $\hat \tau(x,\beta)$ computed from (\ref{tauinlimitalpahtoinfty})
for particles that are released at the upper cylinder surface at radial position $x$ for various $\beta$. We further compare the analytic computations with results from Brownian simulations obtained with 10000 particles and $\alpha=50$. (b) Comparison of $\hat \tau(0,\beta)$ with the approximation $\hat \tau^{(0)}(0,\beta)_{ap} \approx {a_0(\beta)}/{I_0(\frac{\pi}{2\beta})}$ given in (\ref{approxtauuppersurf}). }
\label{fig_MFPT}
\end{center}
\end{figure}

%
\subsubsection{Impact of truncating the series for $\hat \tau(x,y)$ in(\ref{tauinlimitalpahtoinfty})}
\label{section_truncation}
We now study the error induced by truncating the sum in (\ref{tauinlimitalpahtoinfty}) at levels $n\sim1$ by considering the truncated series
\bea\label{tauinlimitalpahtoinfty_trunc}
\hat \tau^{(n)}(x,y)=\left\{
\begin{array}{l}
\displaystyle  \sum_{i=0}^n  b_i(\beta) \frac{I_0(l_i x)}{I_0(l_i)} v_i(y) \,, \quad x\le 1 \\ \\
\displaystyle \frac{a_0(\beta)}{\sqrt 2} + \sum_{i=1}^n  a_i(\beta) \frac{K_0(k_i x)}{K_0(k_i)} u_i(y)
+ \frac{\ln x}{2\pi \beta } \,, \quad  1 \le x \ll \alpha  \,,
\end{array}\right.
\eea
To evaluate the error induced by the truncation, we first compute the coefficients $a_n$ and $b_n$ with high precision (using a truncation level $n\sim200$), and then use these values in (\ref{tauinlimitalpahtoinfty_trunc}). In Fig.~\ref{fig_MFPT_trunc}, we show the effect of the truncations for various $n$ and $\beta$: interestingly, the numerical analysis reveals that for $\beta \lesssim 1$, truncating at $n=0$ or $n=1$ already provides a very good approximation. The accuracy of the truncation depends on $\beta$, and $n$ has to be increased for larger $\beta$ in order to maintain a similar accuracy (Fig.~\ref{fig_MFPT_trunc}c). In Fig.~\ref{fig_MFPT}a-c, we plot the effect of the truncation as a function of $x$ for $y=\beta$ (particles are released at the upper surface), and in Fig.~\ref{fig_MFPT}d-f, the starting position $y$ is reduced to $y=0.7\beta$, $y=0.4\beta$ and $y=0.1\beta$.

Due to the truncation, at the patching boundary $x=1$, (\ref{tauinlimitalpahtoinfty_trunc}) has a small  discontinuity $\Delta^{(n)}(y)=\hat \tau^{(n)}(1^+,y)-\hat \tau^{(n)}(1^-,y)$. For example, for $n=0$ and $y=\beta$ (see Fig.~\ref{fig_MFPT_trunc}a-c) we obtain $\Delta^{(0)}(\beta) = \frac{a_0(\beta)}{\sqrt{2}} - b_0(\beta) \approx a_0(\beta)(1-\frac{2}{\sqrt{2}})$, where we used $a_0(\beta) \approx b_0(\beta)$  (see Fig.~\ref{fig_a0}c-d).

Finally, for $\beta \lesssim 1$, when a Brownian particle is released at the center of the upper surface $(x=0,y=\beta)$, using the truncation $n=0$, we obtain from (\ref{tauinlimitalpahtoinfty_trunc})
\bea\label{approxtauuppersurf}
\hat \tau(0,\beta) \approx \hat \tau^{(0)}(0,\beta)=\frac{b_0(\beta)}{I_0(\frac{\pi}{2\beta})} \approx \frac{a_0(\beta)}{I_0(\frac{\pi}{2\beta})} = \frac{\sqrt 2}{I_0(\frac{\pi}{2\beta})} \hat \tau(1)\,,
\eea
where we additionally used that $a_0(\beta\approx b_0(\beta)$ and $\hat \tau(1)=a_0/\sqrt 2$. In Fig.~\ref{fig_MFPT}b, we test this approximation as a function of $\beta$ by comparing it with $\hat \tau(0,\beta)$, computed with high accuracy ($n\sim 200$). We find that this approximation is valid until $\beta \sim 1$.  }

\begin{figure}[h!]
\begin{center}
       \subfigure[$\beta=0.5$]{\includegraphics[scale=0.35]{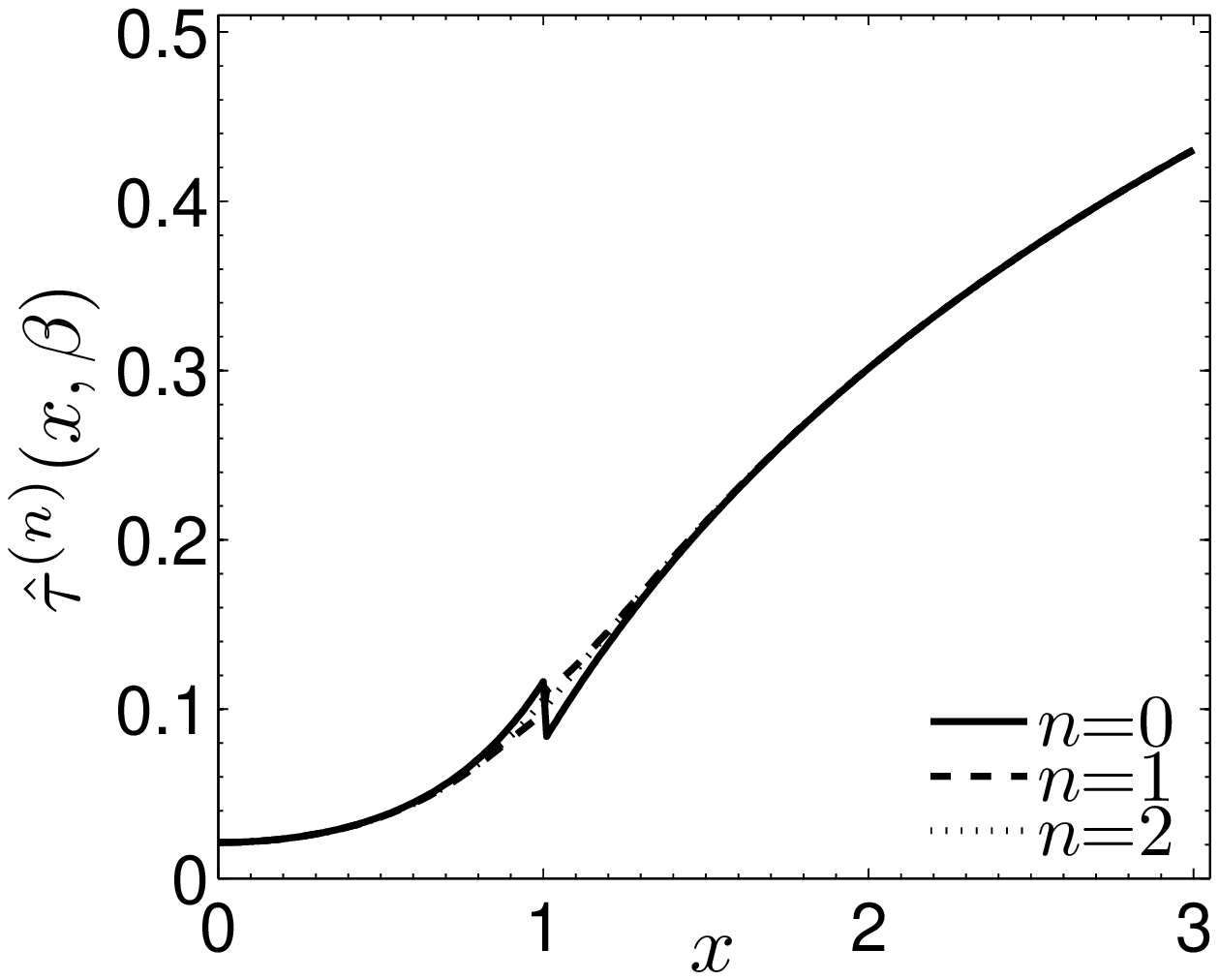}}
       \subfigure[$\beta=1$]{\includegraphics[scale=0.35]{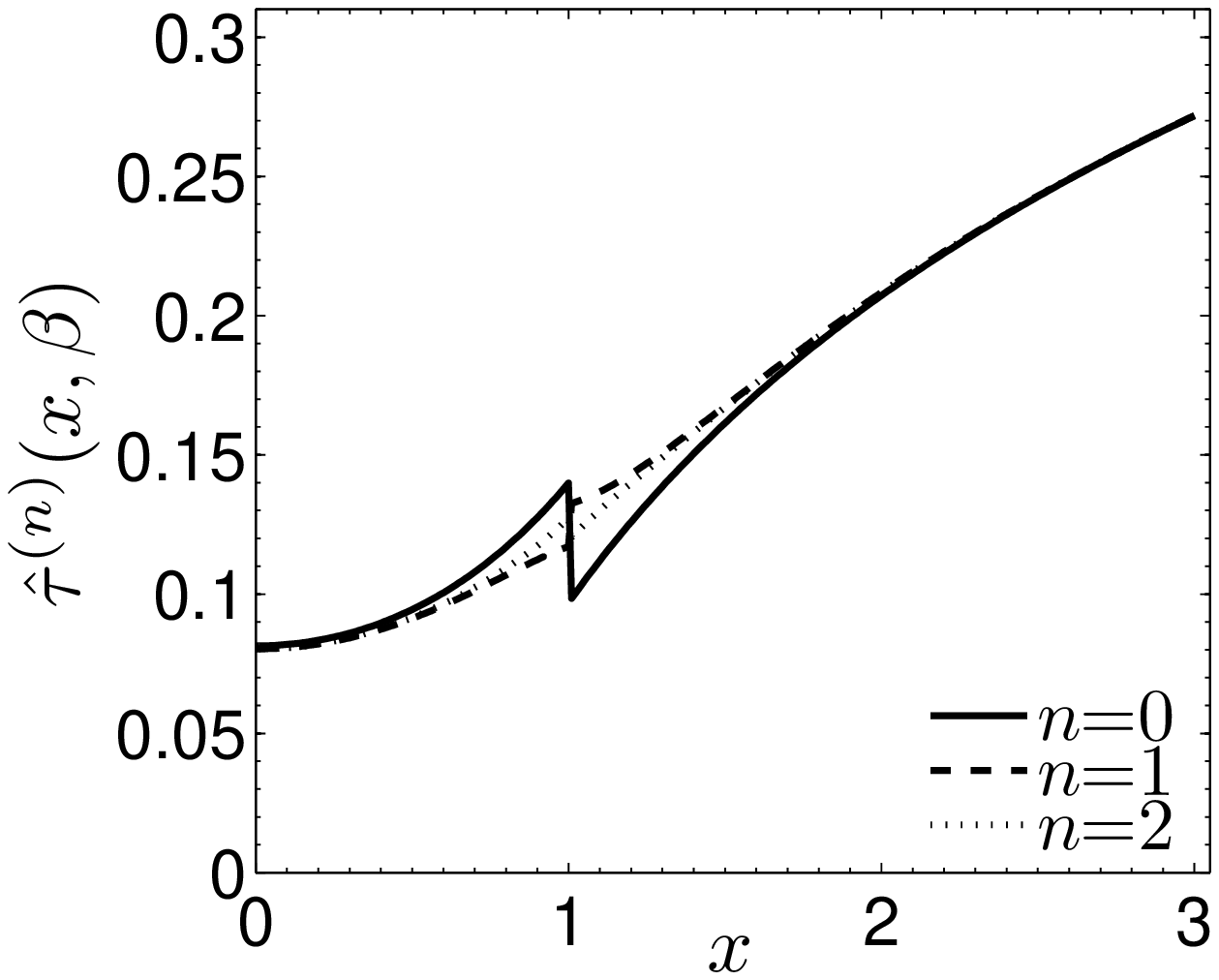}}
       \subfigure[$\beta=10$]{\includegraphics[scale=0.35]{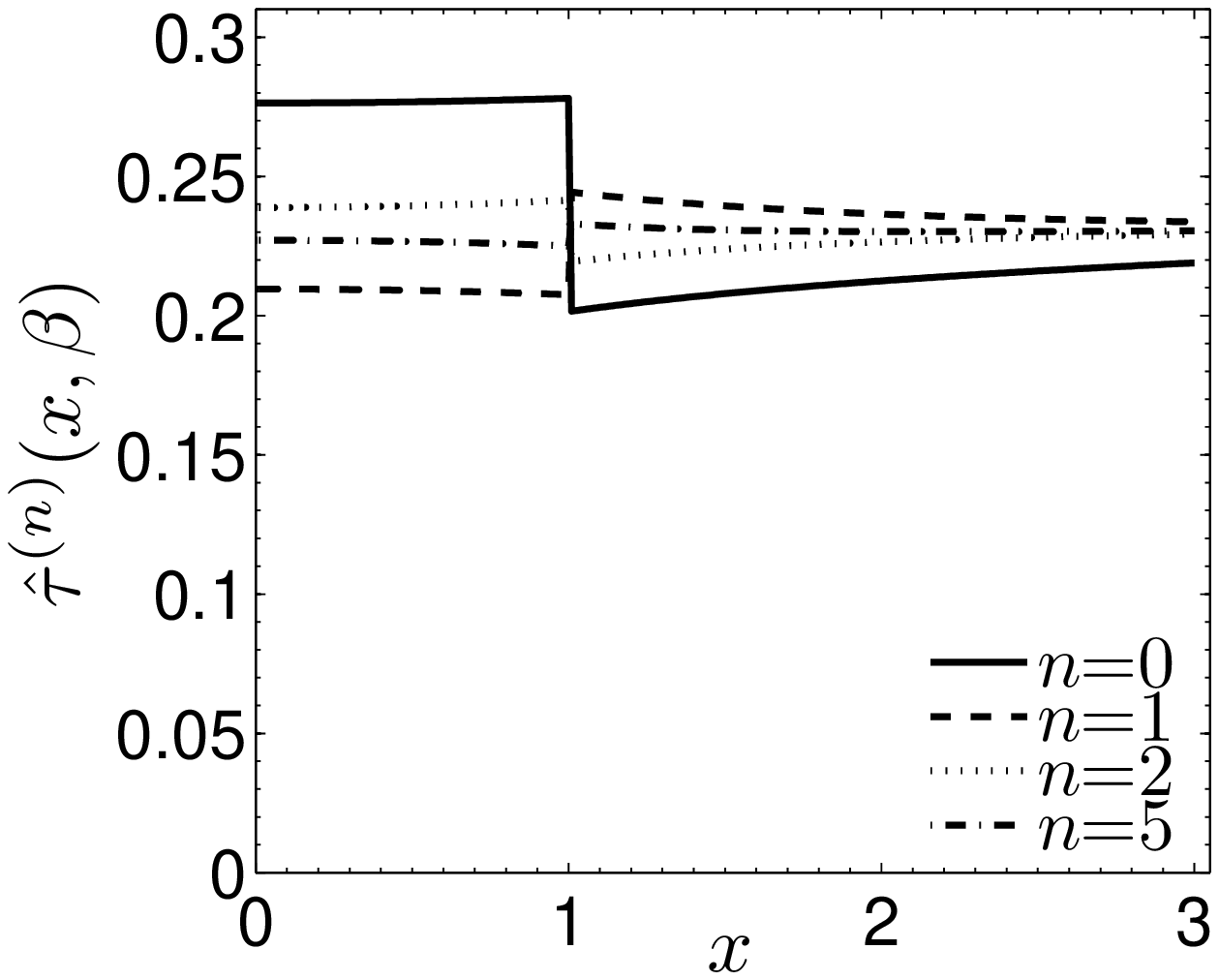}}
       \subfigure[$\beta=1$ and $y=0.7\times \beta$]{\includegraphics[scale=0.35]{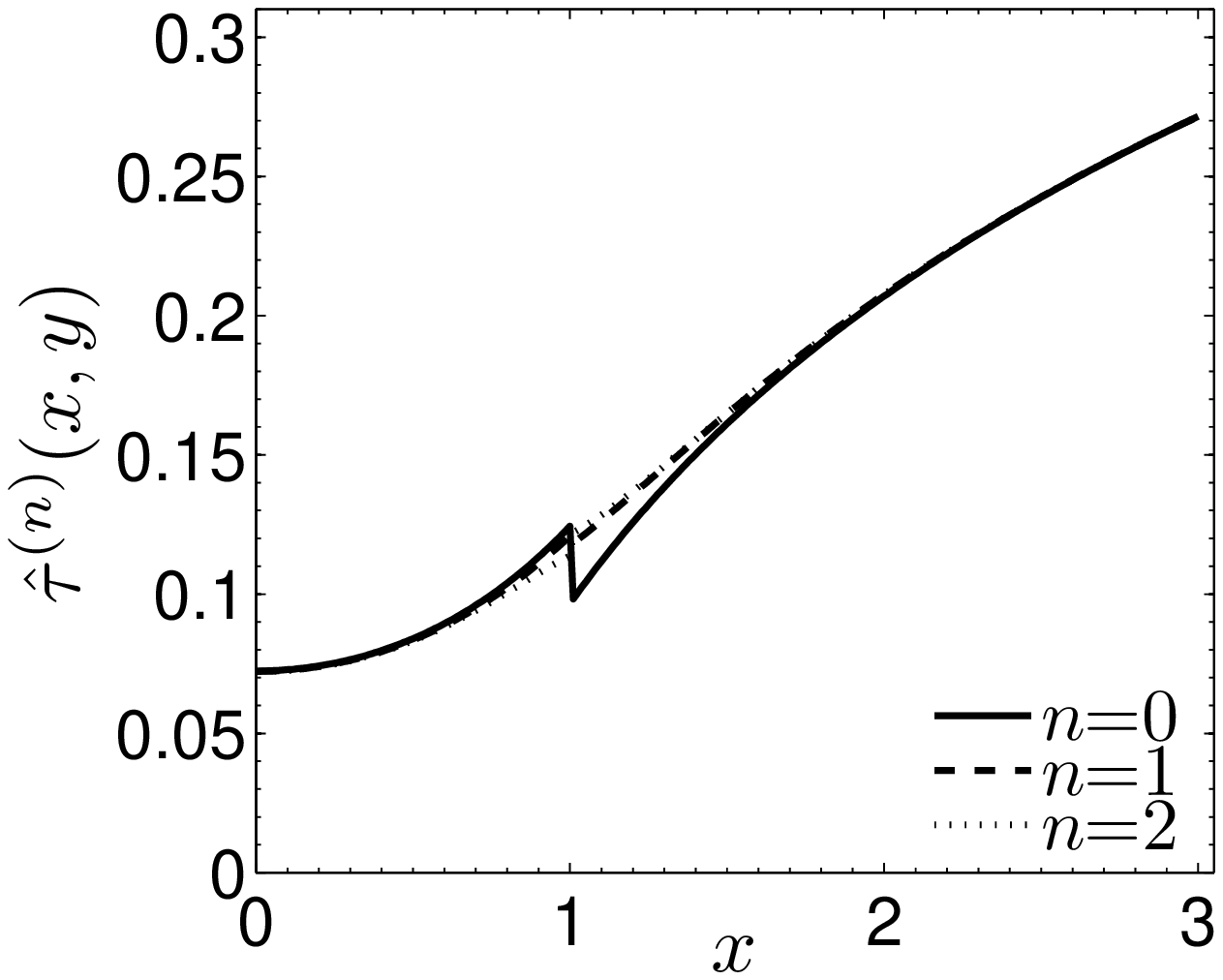}}
       \subfigure[$\beta=1$ and $y=0.4\times \beta$]{\includegraphics[scale=0.35]{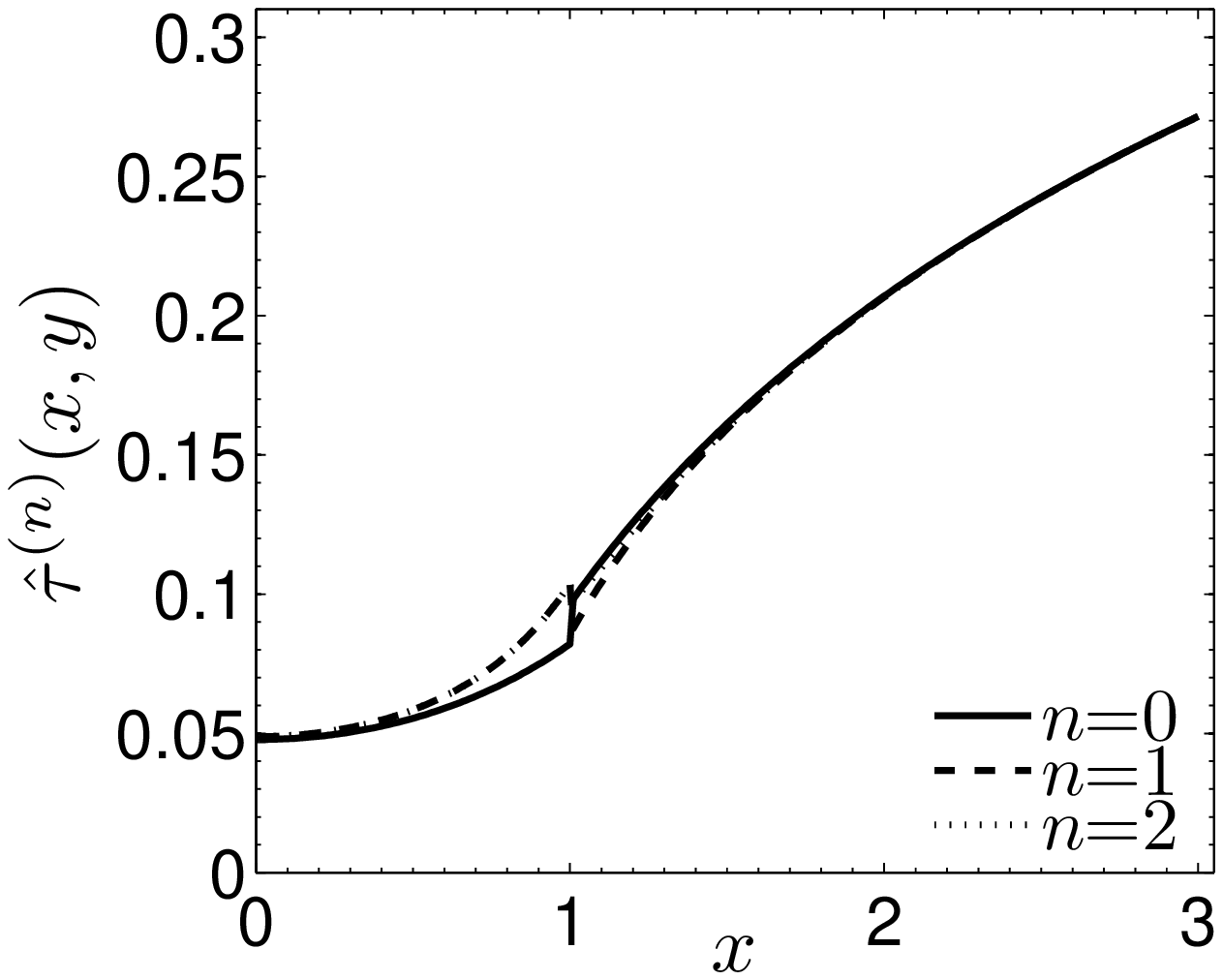}}
       \subfigure[$\beta=1$ and $y=0.1\times \beta$]{\includegraphics[scale=0.35]{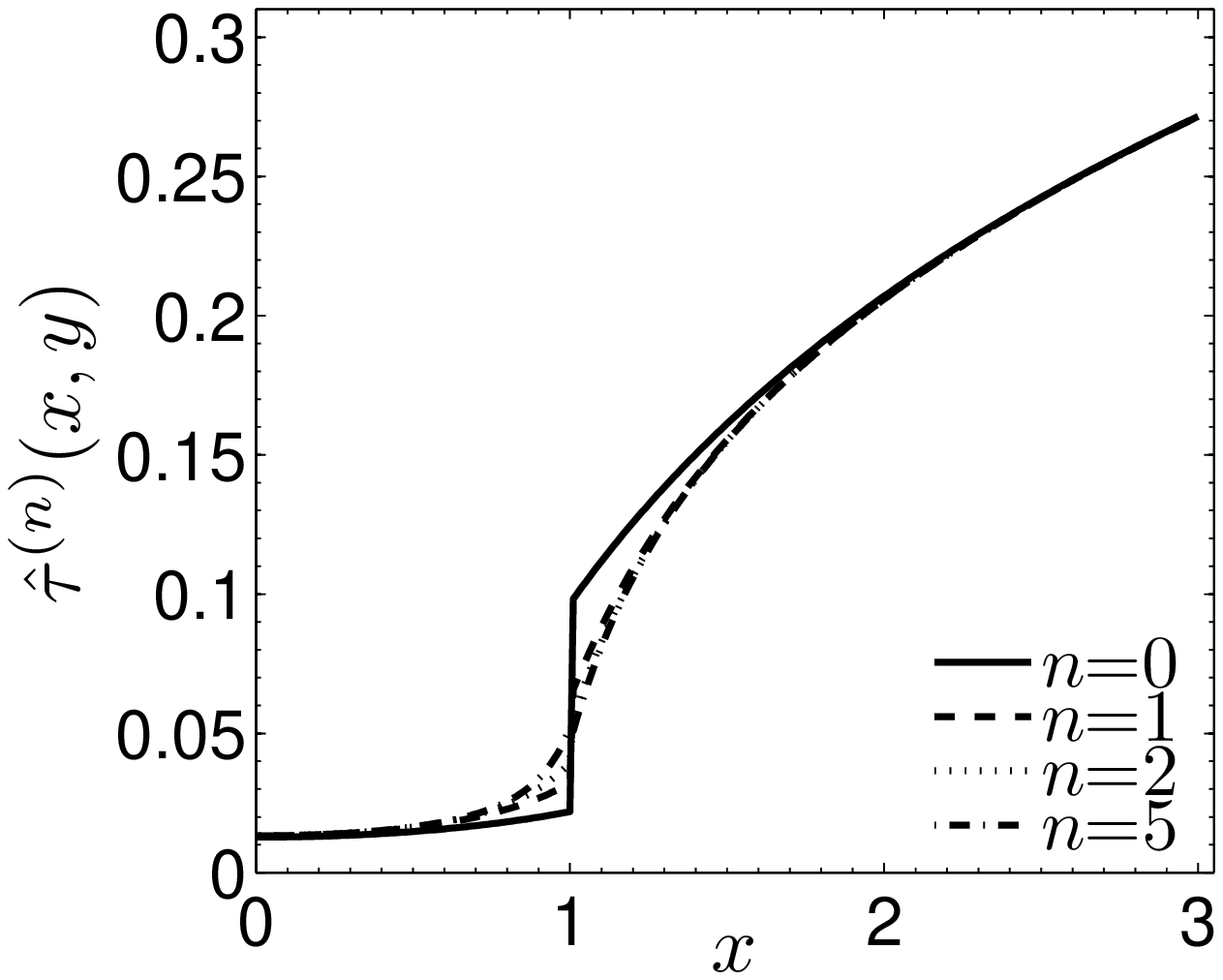}}
       \caption{The truncated NET $\hat \tau^{(n)}(x,y)$ from (\ref{tauinlimitalpahtoinfty_trunc}) for various truncation levels $n$ and several $\beta$. The figure shows that for $\beta \lesssim 1$ very good results  can be obtained by considering a low truncation level $n\sim2$.
       (a)-(c): $\hat \tau^{(n)}(x,\beta)$ for various $\beta$. (d)-(f): $\hat \tau^{(n)}(x,y)$ for $\beta=1$ and various $y$.    }
\label{fig_MFPT_trunc}
\end{center}
\end{figure}

%
\subsubsection{Analogy with the electrified disk problem}
%
For $\beta\gg 1$ and $\alpha \gg1 $, an asymptotic solution for
$\hat \tau(x,y)$ can be obtained by considering the analogy with the electrified
disk problem: the total outflux $-1$ through the hole leads to the electrified disk problem with a disk charge $Q=-1$, and using the capacitance $C=4$ of the unit disk \cite{BookSneddon,BookSmythe}, we find that the disk
potential is ${Q}/{C}=-{1}/{4}$. Using the solution $U(x,y)$ of
the electrified disk problem with disk potential $-1/4$ \cite{BookSneddon}, we obtain the asymptotic correspondence
\bea\label{asymptaulargebeta}
\hat \tau(x,y) = U(x,y) + \frac{1}{4}\,,\quad \alpha \gg1 \,\,, \beta \gg 1\,.
\eea
Hence, for large $\beta$, $U(x,y)$ determines the shape of the boundary layer. Furthermore, by comparing (\ref{asymptaulargebeta}) with (\ref{tauinlimitalpahtoinfty}), we recover that $\frac{a_0}{\sqrt 2}=\frac{1}{4}$ for $\beta\sim \alpha \gg 1$.

%
%
\section{Conditional probability to reach the small target before leaving a laterally open cylinder}
%
When the cylinder is open at the lateral boundary, we shall now compute the conditional probability $p(x,y)$ that a Brownian particle, initially at position $(x,y)$, reaches the small target disk before leaving the cylinder through the lateral opening. Because the geometry of the synaptic cleft can be approximated by a laterally
open cylinder \cite{Barbour_JNeurosc2001,Lisman_JNeurophys2004,TafliaHolcman2010}, we will use our computations to estimate the efficiency of receptor activation at a synapse. Indeed, at the presynaptic site, vesicles release neurotransmitters into the synaptic cleft, and the diffusing neurotransmitter either bind to and thereby activate receptors located on postsynaptic terminal, or they leave the synaptic cleft without activating a receptor. We will first derive a general expression for the conditional probability $p(x,y)$ to hit the small target before exiting, and then compute average values for uniform initial distributions. Finally, we will determine the leading order behavior in a flat cylinder with $R \gg a$ and $h\ll R$.

The conditional probability $p(x,y)$ satisfies the Laplace equation \cite{BookSchuss,BookKarlinTaylor1}
\bea
\begin{array}{c}
\ds \( \frac{1}{x} \frac{\p}{\p x} x\frac{\p}{\p x} +
\frac{\p^2}{\p y^2}\) p(x,y) = 0\,,
\quad x\in \Omega   \label{scaledEqPi}\\
\ds p(x,0)= 1\,\mbox{ for } x < 1 \,,  \quad p(x,\alpha) = 0 \\
\ds \frac{\partial}{\partial y} p(x,0) = 0 \mbox{ for }  1 < x < \alpha \,, \quad
\ds \frac{\partial}{\partial y} p(x,\beta)= 0 \,.
\end{array}
\eea
Similarly to the analysis of $\hat \tau(x,y)$ in the previous section, we solve
(\ref{scaledEqtau_2}) in the subdomains $\Omega_i$ and $\Omega_o$, and then patch the two solutions $p_i(x,y)$ and $p_o(x,y)$ at the boundary $x=1$. The general expressions are
\bea\label{solutionPi}
p(x,y)=\left\{
\begin{array}{l}
\displaystyle  1 - \sum_{n=0}^\infty  b_n^p \frac{I_0(l_n x)}{I_0(l_n)} v_n(y) \,, \quad x\le 1 \\ \\
\displaystyle  \(1 - \frac{a_0^p}{\sqrt 2} \)\frac{\ln \( \frac{\alpha}{x}\) }{\ln \alpha} -
\sum_{n=1}^\infty  a_n^p \frac{G_0(k_n x,k_n \alpha)}{G_0(k_n,k_n \alpha)} u_n(y)
\,, \quad 1 \le x \le  \alpha
\end{array}\right.
\eea
where
\bea
G_0(x,y) = I_0(x)K_0(y) - K_0(x) I_0(y),\nn
\eea
and the unknown coefficients $a_n^p$ and $b_n^p$ are functions of $\alpha$ and
$\beta$ and are related by a relation similar to (\ref{an_bn}).
The coefficient $a_n^p$ resp. $b_n^p$ are given by
\bea\label{prob_eqforan_dim3}
\begin{array}{lll}
\ds \sum_{m=0}^\infty (\beta_n + \alpha_m^p ) \xi_{nm} a_m^p &=&  \ds \xi_{n0} \gamma_0^p \\
\ds \sum_{m=0}^\infty (\beta_m + \alpha_n^p) \xi_{mn} b_m^p &=&  \ds \gamma_0^p \delta_{n0} \,,
\end{array}
\eea
with
\bea\label{prob_coeffalphabetagamma_dim3}
\alpha_0^p=\frac{1}{\ln \alpha}\,,\,\, \alpha_n^p =
-k_n \frac{G_1(k_n,k_n \alpha)}{G_0(k_n,k_n \alpha)}\,(n\ge 1)\,,  \quad
\beta_n = l_n \frac{I_1(l_n)}{I_0(l_n)} \,, \quad  \gamma_0^p = \frac{\sqrt 2}{ \ln \alpha}\,,
\eea
and
\bea
G_1(x,y)=\frac{\p}{\p x} G_0(x,y) = I_1(x)K_0(y) + K_1(x) I_0(y)\,. \nn
\eea
For the $\beta_n$ we omitted the superscript $^p$ because they coincide with the $\beta_n$ already defined in (\ref{defbetan}).

%
\subsection{Conditional probabilities with uniform initial distributions}
%
The fraction $p(x)$ of Brownian particles that eventually reach the target starting initially uniformly distributed at $x$ is
\bea
p(x) = \frac{1}{\beta}\int_0^\beta p(x,y)dy = \left\{
\begin{array}{l}
\displaystyle  1 - \sum_{n=0}^\infty  \frac{b_n^p}{\beta l_n} \frac{I_0(l_n x)}{I_0(l_n)}  \,,\quad   x\le 1 \\ \\
\displaystyle \( 1 - \frac{a_0^p}{\sqrt 2}\) \frac{\ln\(\frac{\alpha}{x}\)}{\ln \alpha} =
p(1) \frac{\ln\(\frac{\alpha}{x}\)}{\ln \alpha}
\,, \quad 1 \le x \le  \alpha
\end{array}\right. \label{prob_pi(x)}
\eea
where $p(1)=1 - {a_0^p}/{\sqrt 2 }$. For particles starting initially uniformly distributed in $\Omega$, the average probability is
\bea
p = \frac{2}{\alpha^2}
\int_0^\alpha p(x) x dx = p(1)
\frac{\alpha^2-2\ln\alpha -1}{2 \alpha^2\ln\alpha} + \frac{1}{\alpha^2} -
\frac{2}{\beta \alpha^2} \sum_{n=0}^\infty  \frac{b_n}{l_n^2}
\frac{I_1(l_n)}{I_0(l_n)}\,. \label{prob_pi_dim3}
\eea
When the particles are initially uniformly released at the upper surface
within an area $x\le x_0$, the fraction $p_\beta(x_0)$ that will
reach the target before escaping through the lateral opening is
\bea\label{funcprobbetax0}
p_\beta(x_0) = \frac{2}{x_0^2} \int_0^{x_0} p(x,\beta) x dx =
\left\{
\begin{array}{l}
\displaystyle  1 - \frac{2}{x_0}\sum_{n=0}^\infty  (-1)^n \frac{b_n^p \beta_n}{l_n^2} \frac{I_1(l_n x_0)}{I_1(l_n)}   \,, \quad x_0 \le 1 \\ \\ \\
\ds \frac{ p_\beta(1)}{x_0^2} + \frac{2}{x_0^2}  \(1 - \frac{a_0^p}{\sqrt 2} \) \frac{x_0^2\(1 + 2\ln(\frac{\alpha}{x_0})\) - (1 +2 \ln \alpha) }{4 \ln \alpha } \\
\ds +\frac{2}{x_0^2} \sum_{n=1}^\infty (-1)^n  \frac{a_n^p \alpha_n^p }{k_n^2} \( \frac{x_0 G_1(k_n x_0,k_n \alpha)}{G_1(k_n,k_n \alpha)} -1 \) \,,  \quad  1 \le x_0 \le \alpha \,.
\end{array}\right.
\eea
%
\subsection{Asymptotic expressions for the conditional probability in a flat cylinder with $\alpha \gg 1$ and $\beta \ll \alpha$}
%
To obtain an asymptotic expression for the conditional probability $p$ in the limit $\alpha \gg 1$ and $\beta \ll \alpha$, we estimate $a_n^p$ and $b_n^p$. For $\alpha \gg 1$ and $\beta/\alpha\ll 1$ we have
\bea
\alpha_0^p=\frac{1}{\ln \alpha}\,, \quad \alpha_n^p \approx k_n \frac{K_1(k_n)}{K_0(k_n)}= \alpha_n, \,\,(n\ge 1) \,, \nn
\eea
where $\alpha_n$ are given in (\ref{asympalphanbetantaulargealpha}). For the scaled coefficients $\tilde a_n^p = \frac{\ln \alpha} {2\pi \beta} a_n^p$ we obtain from (\ref{prob_eqforan_dim3}) the asymptotic equations
\bea\label{eq_prob_antilde}
(\beta_n + \frac{1}{\ln \alpha}) \xi_{n0} \tilde a_0^p + \sum_{m=1}^\infty (\beta_n + \alpha_m ) \xi_{nm} \tilde a_m^p = \frac{1}{\sqrt{2} \pi \beta} \xi_{n0}\,.
\eea
When $\alpha$ is large such that $\frac{1}{\ln\alpha}\ll \beta_n$, (\ref{eq_prob_antilde}) reduces to (\ref{equa_n}) for the coefficients $a_n$ of $\hat \tau(x,y)$, hence, $\tilde a_n^p \approx a_n$. Because $\beta_n$ are monotonically increasing with $n$, if $\frac{1}{\ln\alpha} \ll \beta_n$ for $n=0$, this is also valid for $n>0$. Hence, by setting $n=0$, we obtain the condition $\frac{1}{\ln\alpha} \ll
\frac{\pi}{2\beta} \frac{I_1(\frac{\pi}{2\beta})}{I_0(\frac{\pi}{2\beta})}$, satisfied for
small $\beta$ when $\ln\alpha \gg \beta$, and for large $\beta$ when $\ln\alpha \gg
\beta^2$. In this limit, the asymptotic expressions for $a_n^p$ and $b_n^p$ are
\bea\label{anpbnp}
a_n^p=\frac{2\pi \beta}{\ln \alpha} a_n\quad  \mbox{and} \quad b_n^p=\frac{2\pi \beta}{\ln \alpha} b_n\,.
\eea
Finally, by inserting $a_n^p$ and $b_n^p$ into (\ref{solutionPi}) we obtain for for $\ln \alpha \gg \frac{2\beta}{\pi} \frac{I_0(\frac{\pi}{2\beta})}{I_1(\frac{\pi}{2\beta})}$ the asymptotic formula
\bea\label{p(x,y)largealpha}
p(x,y)=\left\{
\begin{array}{l}
\displaystyle  1 - \frac{2\pi \beta}{\ln \alpha}  \sum_{n=0}^\infty  b_n \frac{I_0(l_n x)}{I_0(l_n)} v_n(y) \,, \quad x\le 1 \\ \\
\displaystyle  p(1) \frac{\ln \( \frac{\alpha}{x}\) }{\ln \alpha} -
\frac{2\pi \beta}{\ln \alpha}  \sum_{n=1}^\infty  a_n \frac{K_0(k_n x)}{K_0(k_n)} u_n(y)
\,, \quad 1 \le x \ll \alpha
\end{array}\right.
\eea
where $p(1)= 1 - \frac{a_0^p}{\sqrt 2} = 1 - \frac{2\pi \beta}{\ln
\alpha}\frac{a_0}{\sqrt 2}$ (\ref{prob_pi(x)}) and we used
\bea
\frac{G_0(k_n x,k_n \alpha)}{G_0(k_n,k_n \alpha)} \approx \frac{K_0(k_n x)}{K_0(k_n)}\,, \quad \alpha \gg 1 \mbox{ and } x \ll \alpha \,. \nn
\eea
By considering (\ref{tauinlimitalpahtoinfty}) for $\hat \tau(x,y)$, this can be rewritten as
\bea\label{p(x,y)largealpha_tau}
p(x,y)=\left\{
\begin{array}{l}
\displaystyle  1 - \frac{2\pi \beta}{\ln \alpha}   \hat \tau(x,y)\,, \quad x\le 1 \\ \\
\displaystyle 1 + \frac{2\pi \beta}{\ln \alpha} \( 1- \frac{\ln \( \frac{\alpha}{x}\) }{\ln \alpha} \) \hat \tau(1) - \frac{2\pi \beta}{\ln \alpha} \hat \tau(x,y)\,,  \quad 1 \le x \ll \alpha
\end{array}\right.
\eea
showing an interesting relation between the conditional probability and the MFPT.  As an example, for particles that start uniformly distributed at $x=1$ we have
\bea\label{prob_p1_asymp}
p(1) =  1 -\frac{2\pi \beta}{\ln \alpha} \hat \tau(1) \,,
\eea
and using that $\hat \tau(1)\lesssim  \frac{1}{4}$, we find that the probability $p(1)$ approaches one for large $\alpha$. In contrast, for particles that start initially uniformly distributed in $\Omega$ we obtain from (\ref{prob_pi_dim3})
\bea
p \approx  \frac{p(1)}{2\ln\alpha}\,, \quad \alpha \gg 1\,, \label{prob_pi_largealpha_dim3}
\eea
which shows that $p$ tends to zero for large $\alpha$, in contrast
to $p(1)$.

%
\subsection{Numerical evaluations}
%
In Fig.~\ref{fig_prob_pibetaalpha_dim3}a we plot $p(1)=1-\frac{a_0^p}{\sqrt 2}$ as a function of $\alpha$ for various $\beta$ (we numerically solve (\ref{prob_eqforan_dim3}) with high accuracy up to $n\sim 400$). As $\alpha\to 1$, the probability $p(1)$ tends to zero because the particles start close to the lateral boundary. The asymptotic limit of $p(1)$ for large $\alpha$ is one, as shown in (\ref{prob_p1_asymp}). However, the convergence is only logarithmical and therefore very slow. In Fig.~\ref{fig_prob_pibetaalpha_dim3}b we display $p(1)$ as a function of $\beta$ for various $\alpha$: for a fixed value $\alpha >1$, the asymptotic limit of $p(1)$ for $\beta \to 0$ is one, and zero for $\beta\to \infty$. The limit for $\beta \to 0$ is intuitive, because the particles start next to the target. To obtain the asymptotic limit for $\beta\to \infty$ we consider (\ref{prob_eqforan_dim3}) for $a_n^p$ and (\ref{prob_coeffalphabetagamma_dim3}):
because the coefficients $\alpha_n^p$ ($n\ge 1$) and $\beta_n^p$ ($n\ge 0$) tend to zero for large $\beta$, the non-vanishing part of (\ref{prob_eqforan_dim3}) is $\alpha_0^p \xi_{n0}a_0^p = \gamma_0^p \xi_{n,0}$, from which it follows that the asymptotic limit
of $a_0^p$ for large $\beta$ is $\sqrt 2$, and thus $p(1)=1-{a_0^p}/{\sqrt{2}}$ converges to zero in this limit. Indeed, for fixed $\alpha$ and increasing $\beta$, it becomes less probable that a particle reaches the disk before the lateral boundary.

\begin{figure}[h]
\begin{center}
       \subfigure[]{\includegraphics[scale=0.3]{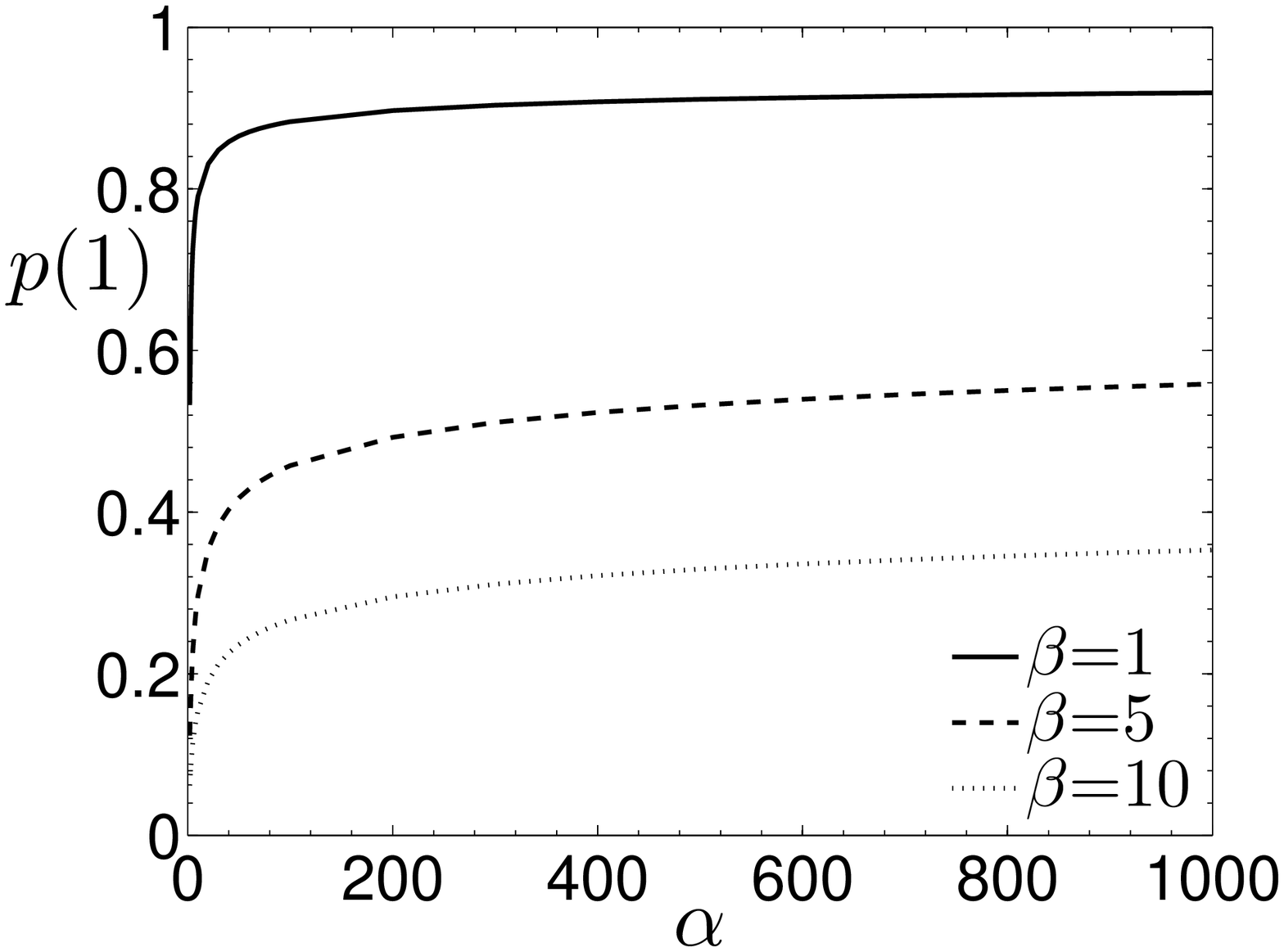}}
       \hspace{0.5cm}
       \subfigure[]{\includegraphics[scale=0.305]{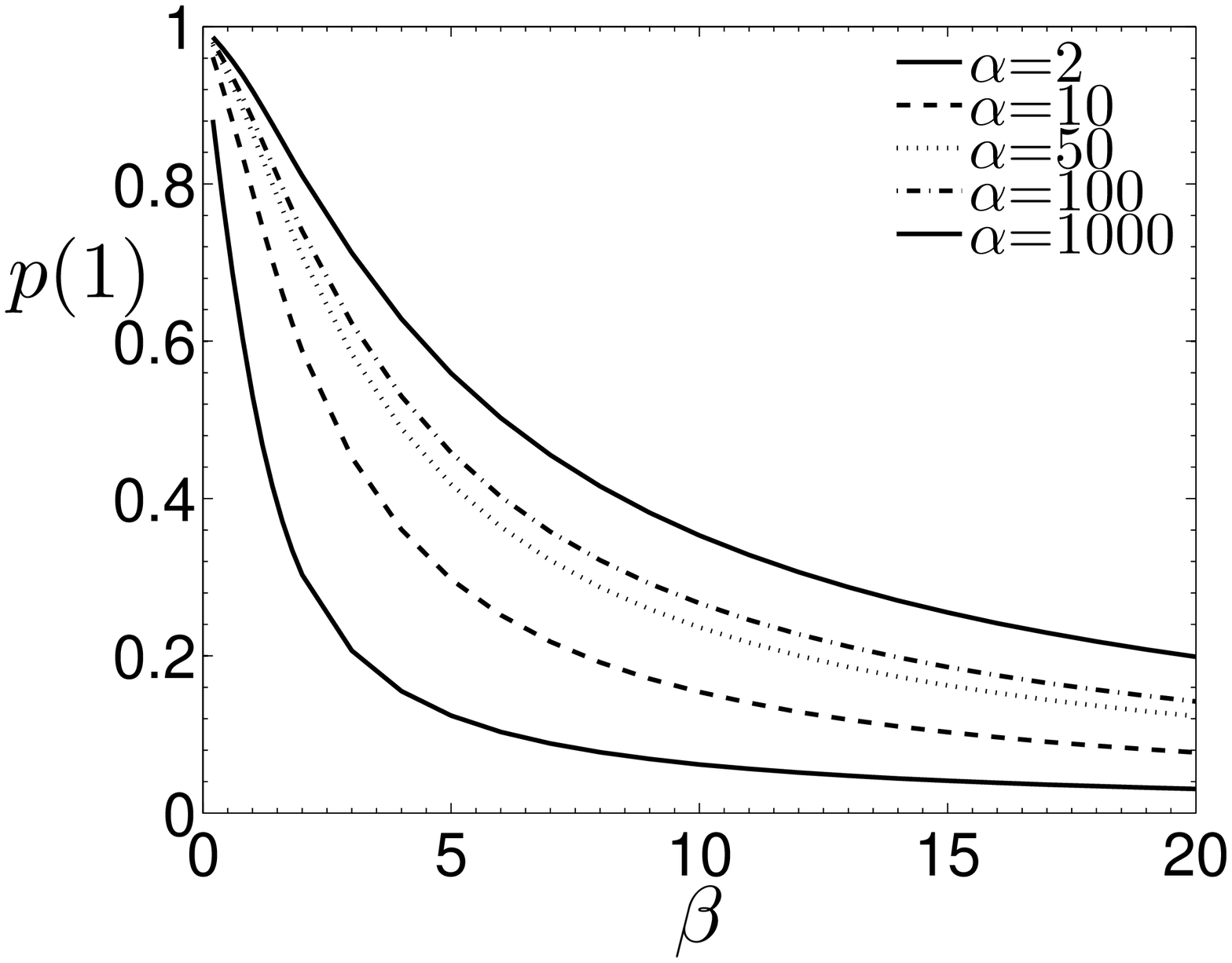}}
  \caption{Average conditional probability $p(1)=1-\frac{a_0^p}{\sqrt 2}$ for Brownian particles starting uniformly distributed at $x=1$ to reach the target before leaving the cylinder through the lateral boundary. The coefficient $a_0^p$ are obtained by truncating and numerically solving (\ref{prob_eqforan_dim3}) with high accuracy.
  (a) $p(1)$ as a function of $\alpha$ for different $\beta$.
  (b) $p(1)$ as a function of $\beta$ for different $\alpha$. }
 \label{fig_prob_pibetaalpha_dim3}
 \end{center}
\end{figure}

We now study the conditional probability for particles that are
released at the upper surface, which is relevant for synaptic transmission. In Fig.~\ref{fig_prob_pi(x,beta)_dim3}a-c we plot $p(x,\beta)$ as a function of $x$ for $\alpha=10$, $\alpha=5$ and $\alpha=100$, and various $\beta$ between 0.1 and 10, and in Fig.~\ref{fig_prob_pi(x,beta)_dim3}a we further show that our analytical computations agree with results from Brownian simulations. In Fig.~\ref{fig_prob_pi(x,beta)_dim3}d we further depict the average probability $p_\beta(x_0)$ when particles are released at the upper surface within an area of radius $x_0$ for similar values $\alpha$ and $\beta$ as in panel (c). In general, Fig.~\ref{fig_prob_pi(x,beta)_dim3} shows that $p(x,\beta)$ is very sensitive to the cylinder height $\beta$ and the release position $x$. For very flat cylinders with $\beta\lesssim 1$, when the particles are released in the area opposite to the hole (for $x<1$), the probability is $\sim 1$, whereas when they are released outside this region (for $x>1$), the probability that they reach the target before exiting decreases
considerably (Fig.~\ref{fig_prob_pi(x,beta)_dim3}a-c). For example, Fig.~\ref{fig_prob_pi(x,beta)_dim3}b, obtained for $\beta=0.5$ and $\alpha=5$, shows that the conditional probability is larger than 90\% when the particles are released at the upper surface within the area $x<1$, whereas it decreases to around 60\% when they are released at $x\sim2$. For cylinder with large $\alpha$ the impact of the release position is much less pronounced (panel (d) with $\alpha=100$). In addition, Fig.~\ref{fig_prob_pi(x,beta)_dim3} shows that the
conditional probability to reach the target is more sensitive to changes in the cylinder height $\beta$ compared to the radius $\alpha$.

\begin{figure}
\begin{center}
\subfigure[$\alpha=10$]{\includegraphics[scale=0.48]{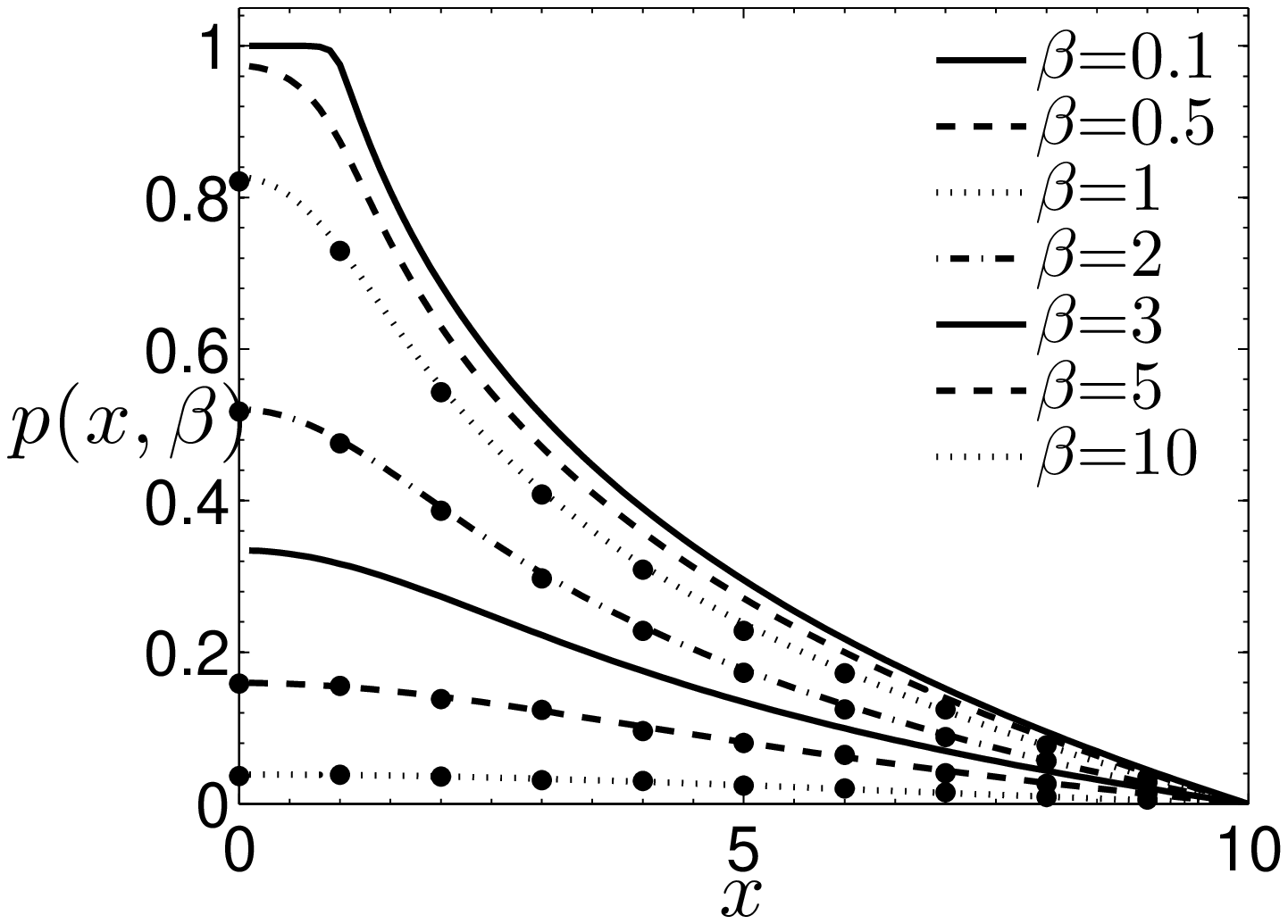}}
\hspace{0.5cm}
\subfigure[$\alpha=5$]{\includegraphics[scale=0.48]{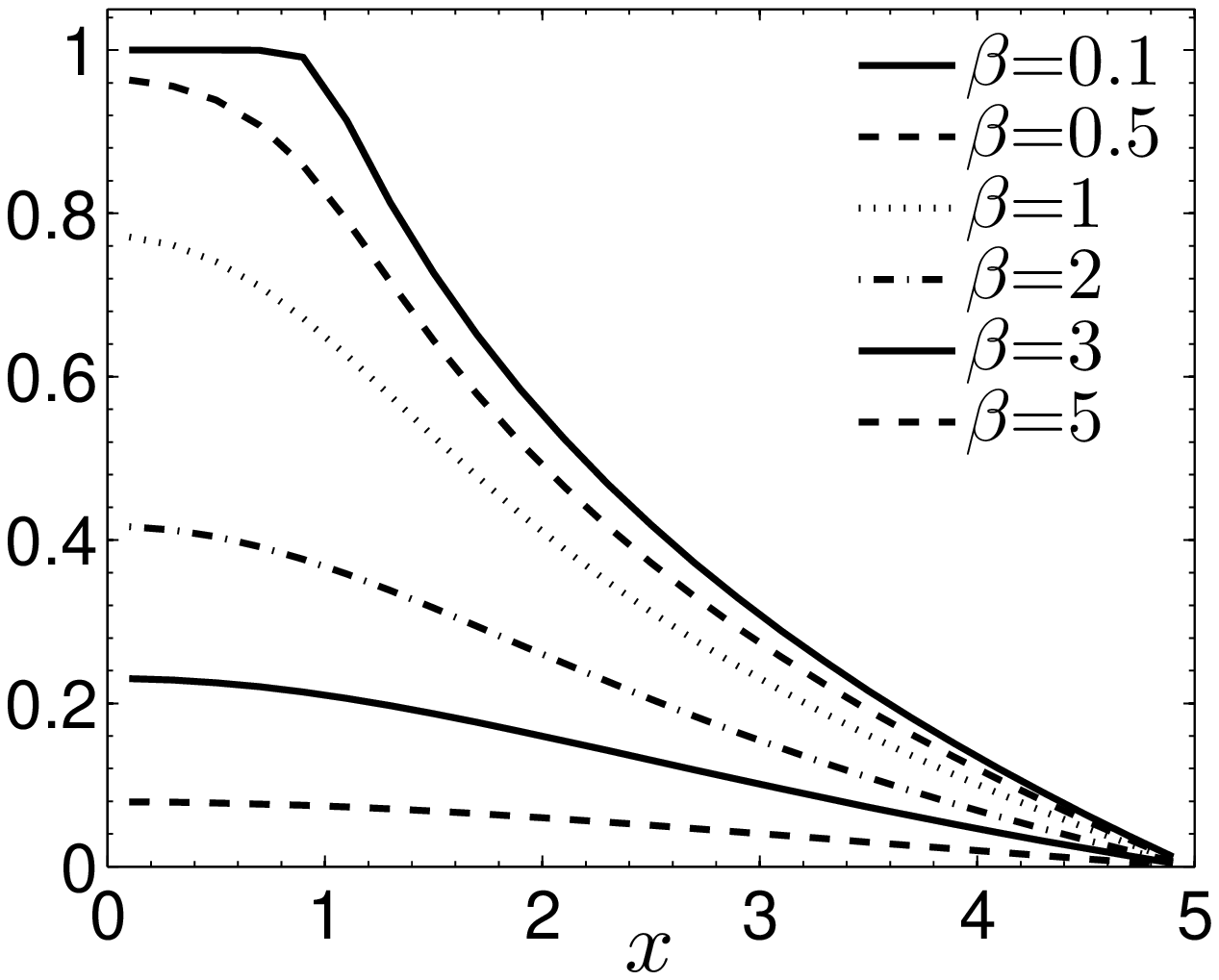}}
\subfigure[$\alpha=100$]{\includegraphics[scale=0.48]{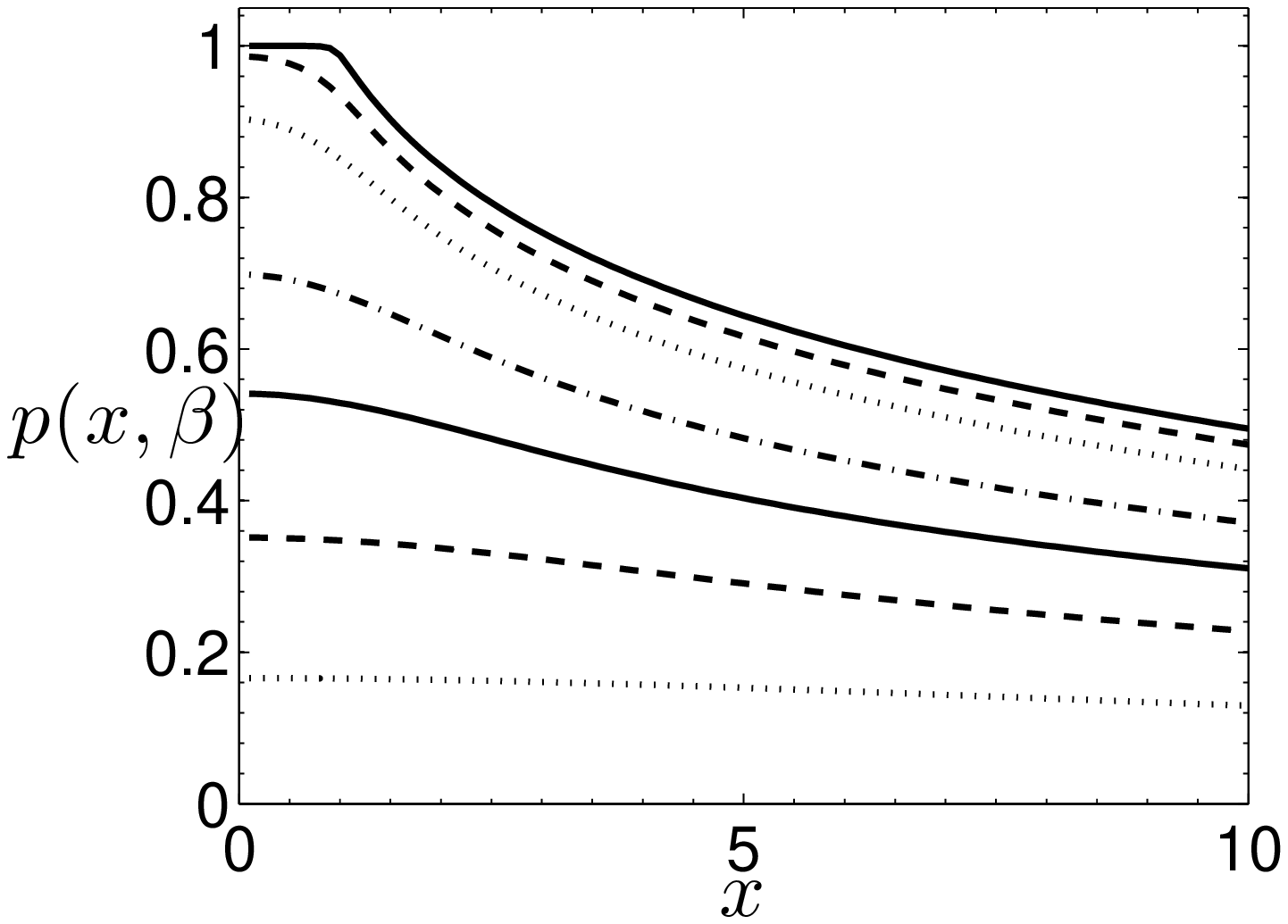}}
\hspace{0.5cm}
\subfigure[$\alpha=100$]{\includegraphics[scale=0.48]{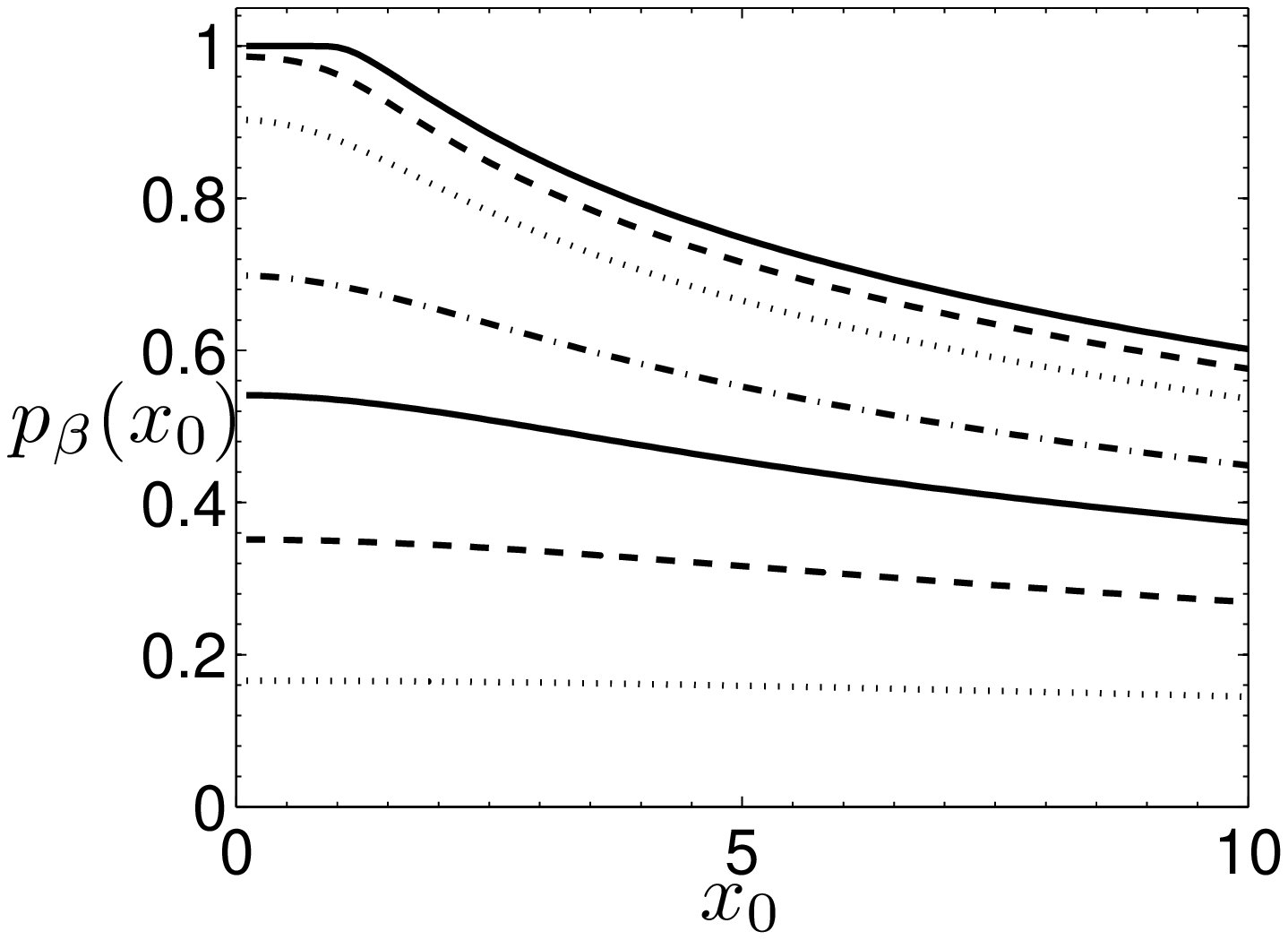}}
  \caption{Panels a-c: Conditional probability $p(x,\beta)$ (from (\ref{solutionPi}))
for a Brownian particle to reach the target at $(x <1,y=0)$  when released at the upper surface at $y=\beta$ and radial position $x$. The data points in (a) are obtained from Brownian simulations with $10^3$ particles. (c) The values for $\beta$  are as in panel (a). (d) Average conditional probability $p_\beta(x_0)$ (from (\ref{funcprobbetax0})) when the particles are released uniformly on the upper surface in an area of radius $x_0$ with values $\beta$ as in (c).}
\label{fig_prob_pi(x,beta)_dim3}
\end{center}
\end{figure}

%
\subsection{Impact of truncating the series for $p(x,y)$ in (\ref{solutionPi})}
\label{section_truncation_prob}
We now proceed similarly to section \ref{section_truncation} and estimate the error induced by truncating the series in (\ref{solutionPi}) at small $n$. We first compute the coefficients $a_i^p$ and $b_i^p$ using (\ref{prob_eqforan_dim3}) with a high accuracy, and then use these coefficients to define the truncated probability
\bea\label{solutionPi_trunc}
p^{(n)}(x,y)=\left\{
\begin{array}{l}
\displaystyle  1 - \sum_{i=0}^n  b_i^p \frac{I_0(l_i x)}{I_0(l_i)} v_i(y) \,, \quad x\le 1 \\ \\
\displaystyle  \(1 - \frac{a_0^p}{\sqrt 2} \)\frac{\ln \( \frac{\alpha}{x}\) }{\ln \alpha} -
\sum_{i=1}^n  a_i^p \frac{G_0(k_i x,k_i \alpha)}{G_0(k_i,k_i \alpha)} u_i(y)
\,, \quad 1 \le x \le  \alpha
\end{array}\right.
\eea
In Fig.~\ref{fig_prob_trunc}, we plot $p^{(n)}(x,\beta)$ for various $\alpha$ and $\beta$ as a function of $x$, and show that truncating at $n=0,1,2$ already gives good approximations (we plot $p^{(n)}(x,\beta)$ for $n=100$ to show the error induced by the low truncations). Similarly as in section \ref{section_truncation}, we conclude that  truncation at $n=1$ or $n=2$ already provides a good approximation for $\beta \lesssim 1$.

\begin{figure}[h!]
\begin{center}
       \subfigure[$\alpha=5$]{\includegraphics[scale=0.52]{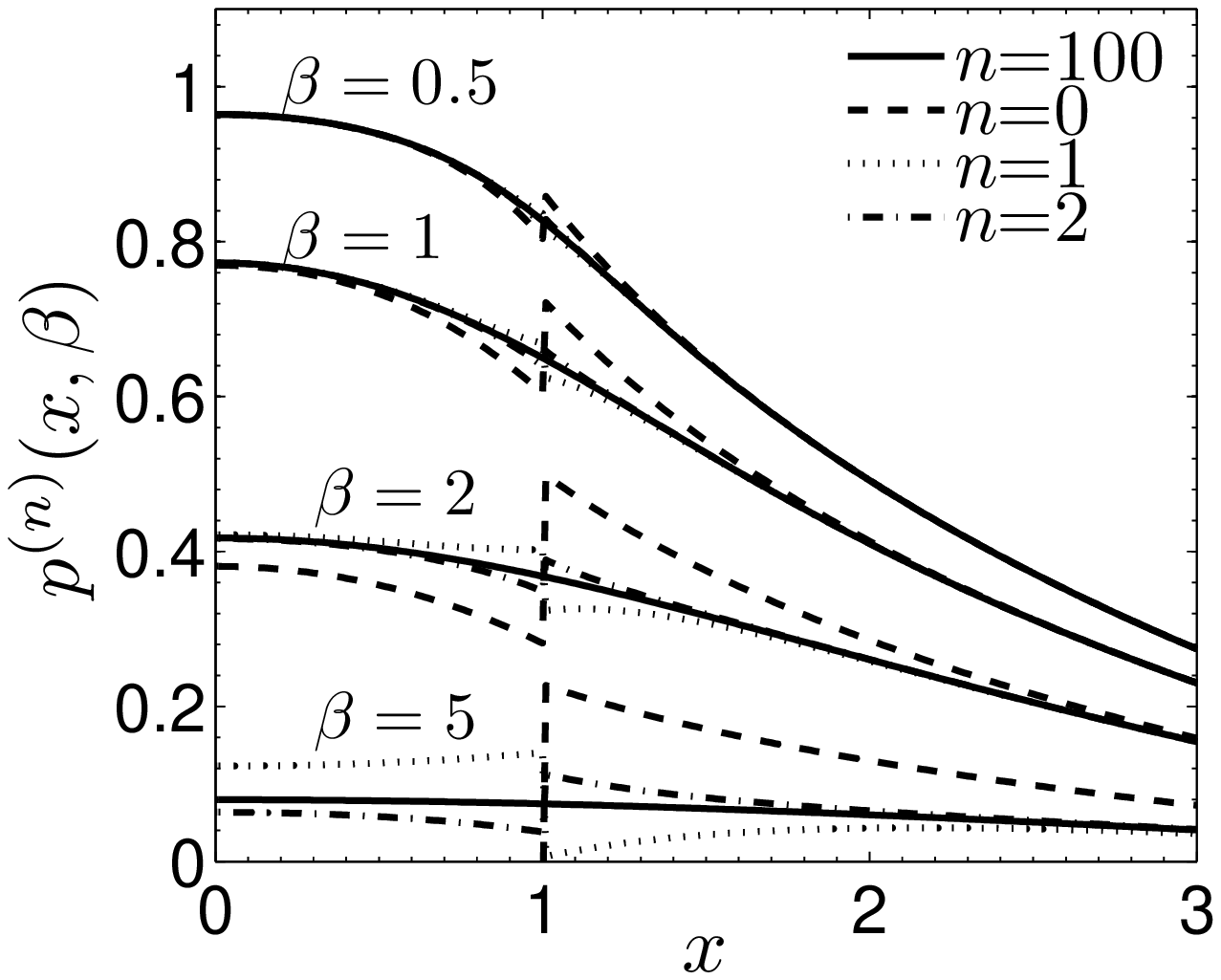}}
       \subfigure[$\alpha=100$]{\includegraphics[scale=0.52]{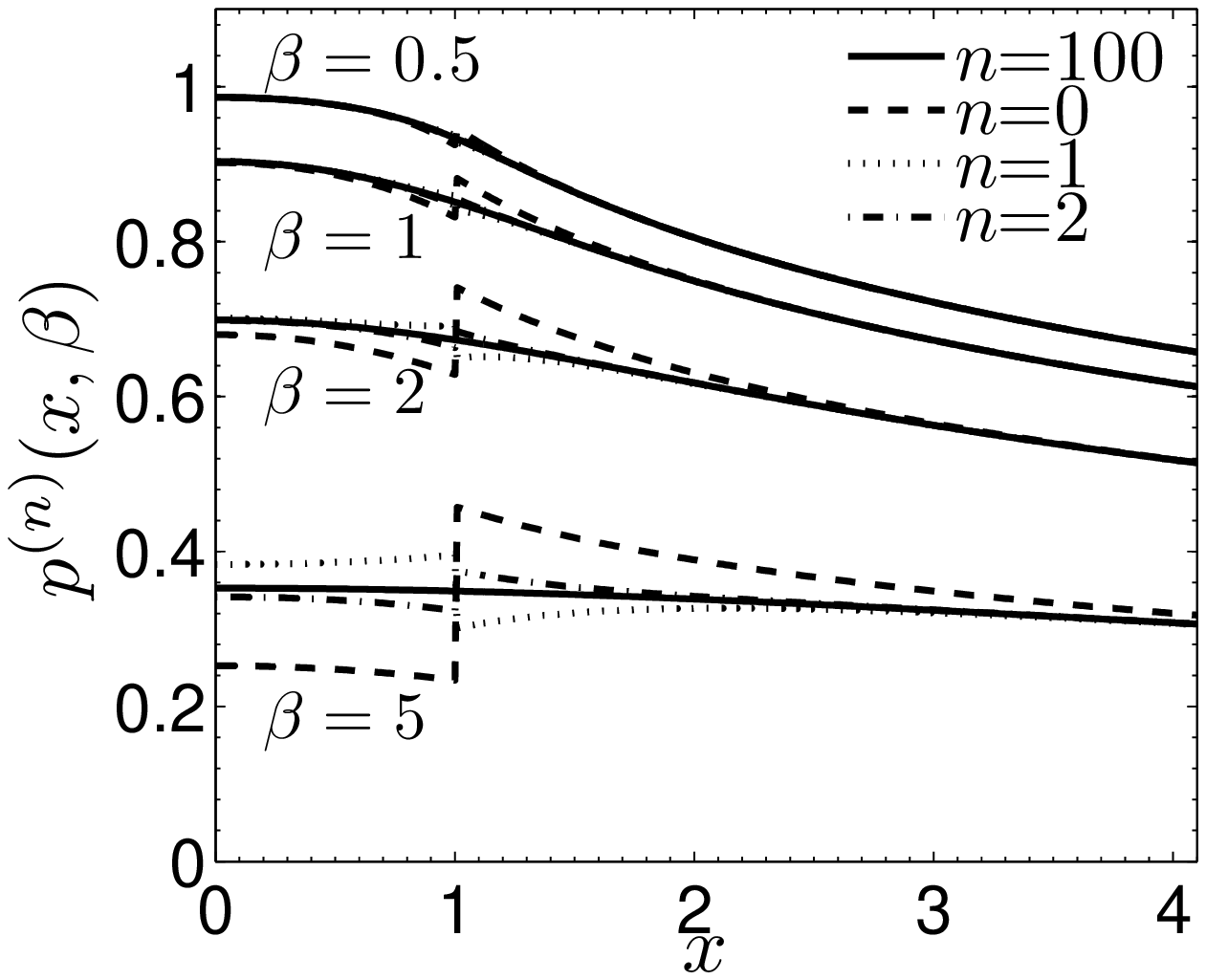}}
       \caption{The truncated conditional probability $p^{(n)}(x,\beta)$ from (\ref{solutionPi_trunc}) for various $n$, $\beta$ and $\alpha$. Similar to Fig.~\ref{fig_MFPT_trunc}, for $\beta \lesssim 1$, we already obtain good approximations for $n\sim2$.   }
\label{fig_prob_trunc}
\end{center}
\end{figure}

%
\section{Conditional mean time to reach a small target before escaping a laterally open cylinder}
%
%
After having estimated the conditional probability $p(x,y)$, we are
now in a position to study the conditional MFPT $\hat \tau_c(x,y)$ to reach a target
before escaping through the lateral cylinder boundary. This analysis will estimate the time scale
of the synaptic response. For example, the conditional time $\hat \tau_c(0,\beta)$ in (\ref{approxtauctau}) provides an estimate for the mean time until postsynaptic receptors become activated after transmitter release into the synaptic cleft at the upper cylinder surface. In a wide cylinder with $\alpha \gg 1$, (\ref{approxtauctau}) shows that the conditional time $\hat \tau_c(0,\beta)$ is roughly by a factor $1/(\ln \alpha)^2$  faster than the mean time $\hat \tau(0,\beta)$ for a laterally closed cylinder, because in a closed cylinder all trajectories that are reflected at the lateral boundary contribute to the mean time, thereby increasing the mean compared to an open cylinder. Thus, to obtain a realistic estimation of the synaptic activation time, we have to consider the conditional time $\tau_c$ for an open cylinder.

To determine the mean conditional time $\hat \tau_c(x,y)$, we use the known conditional probability $p(x,y)$ to solve an equation for the function $A(x,y)=\hat \tau_c(x,y) p(x,y)$, and then we obtain $\hat \tau_c(x,y)$ from $\hat \tau_c(x,y) =\frac{A(x,y)}{p(x,y)}$. The function $A(x,y)$ satisfies \cite{BookGardiner,TafliaHolcman2007,BookKarlinTaylor2}
\bea
\begin{array}{rcl}
\ds \( \frac{1}{x} \frac{\p}{\p x} x\frac{\p}{\p x} +
\frac{\p^2}{\p y^2}\) A(x,y) &=& \ds -\frac{p (x,y)}{|\Omega|}\,,
\quad (x,y) \in \Omega \label{scaledEqtau_A}\\
\ds  A(x,y) &=& 0\,, \quad y=0\,, x<1 \\
\ds  A(x,y) &=&0\,, \quad  x=\alpha  \\
\ds  \frac{\partial}{\partial y}A(x,y)&=&0 \,, \quad  y=0\,, \, x > 1  \\
\ds  \frac{\partial}{\partial y}A(x,y) &=&0\,,  \quad  y=\beta\,.
\end{array}
\eea
Proceeding as in the previous sections, we expand $A(x,y)$ as
\bea
A(x,y)=\left\{
\begin{array}{l}
\ds  \sum_{n=0}^\infty  b^{A}_n \frac{I_0(l_n x)}{I_0(l_n)} v_n(y) + w_i^A(x,y) \,, \quad x\le 1 \\ \\
\ds  \frac{a^{A}_0}{\sqrt2} \frac{\ln \( \frac{\alpha}{x}\) }{\ln \alpha} +  \sum_{n=1}^\infty  a^{A}_n \frac{G_0(k_n x,k_n \alpha)}{G_0(k_n,k_n \alpha)} u_n(y) +
w_o^A(x,y)
\,, \quad 1 \le x \le  \alpha,
\end{array}\right.
\eea
where $w_i^A(x,y)$ and $w_o^A(x,y)$ are the inhomogeneous solutions of (\ref{scaledEqtau_A}) that vanish at $x=1$. The coefficients $a^{A}_n$ and $b^{A}_n$ are related as $a_n$ and $b_n$ in (\ref{an_bn}), and $a^{A}_n$ satisfies
\bea\label{eqforan_A}
\sum_{m=0}^\infty (\beta_n^A + \alpha_m^A ) \xi_{nm} a^A_m = \sum_{m=0}^\infty  \xi_{nm} \gamma_m^A\,,
\eea
where
\bea
\alpha_0^A=\frac{1}{\ln \alpha}\,,\,\, \alpha_n^A =
-k_n \frac{G_1(k_n,k_n \alpha)}{G_0(k_n,k_n \alpha)}\,(n\ge 1)\,,  \quad
\beta_n^A = l_n \frac{I_1(l_n)}{I_0(l_n)} \,,  \label{alphan_A}
\eea
and the $\gamma_n^A$ are implicitly defined through
\bea
\frac{\p}{\p x} \( w_o^A(x,y) -  w_i^A(x,y)\)\Big|_{x=1} = \sum_{n=0}^\infty  \gamma_n^A u_n(y) \,. \label{gamman_A}
\eea
To determine the coefficients $\gamma^A_n$, we first evaluate $w_i^A(x,y)$ and $w_o^A(x,y)$. When $\alpha \gg 1$, $w_i^A(x,y)$ is of the order  $\alpha^{-2}$ and we neglect its contribution in first approximation. Using $p(x,y)$ from (\ref{solutionPi}), the equation for $w_o^A(x,y)$ is
\bea\label{eqfor woA(x,y)}
\begin{array}{rcl}
\ds \( \frac{1}{x} \frac{\p}{\p x} x\frac{\p}{\p x} + \frac{\p^2}{\p y^2}\) w_o^A(x,y)
&=& \ds -\frac{1}{|\Omega|} \( p(1) \frac{\ln\(\frac{\alpha}{x}\)}{\ln \alpha}
- \sum_{n=1}^\infty  a_n^p \frac{G_0(k_n x,k_n\alpha)}{G_0(k_n,k_n\alpha)} u_n(y) \)
\label{eqForWo_A}\\
\ds w_o^A(x,y) &=& \ds 0\,, \quad  x=1 \mbox{ and } x=\alpha \\
\ds \frac{\partial}{\partial y} w_o^A(x,y) &=& \ds 0\,,  \quad  y=\beta \mbox{ and } y=0\,.
\end{array}
\eea
To solve (\ref{eqfor woA(x,y)}) we expand $w_o^A(x,y)$ in terms of $u_n(y)$,
\bea\label{seroeswoA(xy)}
w_o^A(x,y) = w_o^{A(0)}(x) + \sum_{n=1}^\infty w_o^{A(n)}(x) u_n(y)\,,\nn
\eea
and inserting this expansion into (\ref{eqForWo_A}) gives for $w_o^{A(0)}(x)$ the solution
\bea\label{wo[x]_A}
w_o^{A(0)}(x)
= -\frac{p(1)}{4|\Omega|\ln(\alpha) } \( x^2
(\ln\(\frac{\alpha}{x}\) + 1) - (\alpha^2 - \ln\alpha -1) \frac{\ln
x}{\ln \alpha} - (\ln \alpha +1) \)\,.
\eea
The higher order functions $w_o^{A(n)}(x)$ ($n\ge 1$) satisfy the equation
\bea
\begin{array}{c}
\ds \( \frac{1}{x} \frac{\p}{\p x} x\frac{\p}{\p x} - k_n^2\) w_o^{A(n)}(x)
=  \ds \frac{a_n^p}{|\Omega|}\frac{G_0(k_n x,k_n \alpha)}{G_0(k_n,k_n \alpha)}\,, \quad 1 < x < \alpha \\
\ds w_o^{A(n)}(1)= 0\,, \quad  \ds  w_o^{A(n)}(\alpha) =0\,.\nn
\end{array}
\eea
To proceed, we now truncate the series for $w_o^A(x,y)$ in (\ref{seroeswoA(xy)}) at $n=0$ and use only the first order approximation $w_o^A(x,y)\approx w_o^{A(0)}(x)$. We expect that this already provides a good approximation because the coefficients $a_n^p$ are small for for $n\ge 1$ and large $\alpha$, and, as shown in section \ref{section_truncation_prob}, truncation at $n=0$ already gives a very good approximation for $p(x,y$ when $\beta \lesssim 1$. Hence, we expect that our analysis is a valid approximation for large $\alpha$ and small $\beta \sim 1$. With truncation at $n=0$, the parameters $\gamma_n^A$ defined in (\ref{gamman_A}) are
\bea
\gamma_n^A= \gamma_0^A \delta_{n0}\,, \quad \mbox{with} \quad \gamma_0^A =  p(1) \frac{\alpha^2 - 2 (\ln
\alpha)^2  - 2 \ln \alpha-1}{2\sqrt 2 \pi \beta \alpha^2 (\ln \alpha)^2}\approx
\frac{1}{\sqrt 2 \pi \beta} \frac{p(1)}{2 (\ln \alpha)^2} \,.
\eea
Similar to (\ref{anpbnp}), for $\ds{\ln \alpha \gg \frac{2\beta}{\pi} \frac{I_0(\frac{\pi}{2\beta})}{I_1(\frac{\pi}{2\beta})}}$, the asymptotic solutions for $a_n^A$ and $b_n^A$ in terms of $a_n $ and $b_n$ are
\bea\label{anAbnA}
a_n^A=\frac{p(1)}{2 (\ln \alpha)^2} a_n\quad  \mbox{and} \quad b_n^A=\frac{p(1)}{2 (\ln \alpha)^2} b_n\,.
\eea
Finally, we obtain for $\hat \tau_c(x,y)$ the approximation
\bea\label{expr_condMFPT}
\hat \tau_c(x,y) = \frac{A(x,y)}{p(x,y)}=\frac{1}{p(x,y)} \left\{
\begin{array}{l}
\displaystyle   \sum_{n=0}^\infty  b_n^A \frac{I_0(l_n x)}{I_0(l_n)} v_n(y) = \frac{p(1)}{2 (\ln \alpha)^2} \hat \tau(x,y) \,, \quad x\le 1 \\ \\
\displaystyle  \frac{a^{A}_0}{\sqrt2} \frac{\ln \( \frac{\alpha}{x}\) }{\ln \alpha} +  \sum_{n=1}^\infty  a_n^A \frac{K_0(k_n x)}{K_0(k_n)} u_n(y) +
w_o^{A(0)}(x) \,, \quad 1 \le x \ll  \alpha
\end{array}\right.
\eea
where $p(x,y)$ is given in (\ref{p(x,y)largealpha}). To estimate how much $\hat \tau_c(x,y)$ is faster compared to $\hat \tau(x,y)$, we consider particles that are released centrally on the upper surface at position $(0,\beta)$. By taking into account (\ref{p(x,y)largealpha_tau}) for $p(x,y)$, we obtain
\bea
\hat \tau_c(0,\beta) = \frac{1 - \frac{2\pi \beta}{\ln \alpha} \hat \tau(1)}{ 1 - \frac{2\pi \beta}{\ln \alpha} \hat \tau(0,\beta)} \frac{ \hat \tau(0,\beta)}{2 (\ln \alpha)^2}\,.
\eea
Furthermore, using from (\ref{approxtauuppersurf}) the approximation $\hat \tau(0,\beta)\approx \frac{a_0(\beta)}{I_0(\frac{\pi}{2\beta})}= \frac{\sqrt 2}{I_0(\frac{\pi}{2\beta})} \hat \tau(1)$, we obtain for $\beta \lesssim 1$ the approximation
\bea \label{approxtauctau}
\hat \tau_c(0,\beta) \approx \frac{1 - \frac{\sqrt 2\pi \beta}{\ln \alpha} I_0(\frac{\pi}{2\beta}) \hat \tau(0,\beta)}{ 1 - \frac{2\pi \beta}{\ln \alpha} \hat \tau(0,\beta)} \frac{ \hat \tau(0,\beta)}{2 (\ln \alpha)^2}\,.
\eea
In Fig.~\ref{fig_condMFPT} we plot a $\hat \tau_c(x,\beta)$ as a function of $x$ for various $\beta$, and we confirm that our analysis agrees well with results from Brownian simulations. Comparing Fig.~\ref{fig_condMFPT} with Fig.~\ref{fig_MFPT} for $\hat \tau (x,y)$ shows that $\hat \tau_c(x,\beta)$ is roughly by a factor $(2\ln \alpha)^2$ faster compared to $\hat \tau (x,y)$.

\begin{figure}[h!]
\begin{center}
      \includegraphics[scale=0.35]{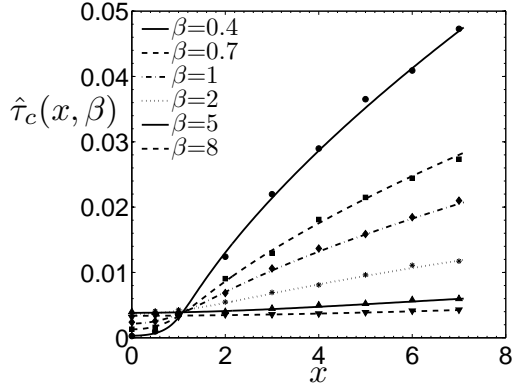}
       \caption{
       Conditional MFPT $\hat \tau_c(x,\beta)$ to reach the hole in a laterally open cylinder for particles that are released on the surface opposed to the hole. The numerical values for $\hat \tau_c(x,\beta)$ are calculated using (\ref{expr_condMFPT}) and $\alpha=50$. The data points are the results of Brownian simulations with 10000 particles and $\alpha=50$.}
       \label{fig_condMFPT}
\end{center}
\end{figure}

%
%
\section{Discussion and conclusion}
%
%
We generalized here the narrow escape problem to a degenerated geometry
defined by a flat cylindrical compartment of height $h$ and radius $R$,
where the absorbing hole is a small circular disk of radius $a$ located centrally on the lower surface.
We analyzed the problem for a laterally closed and open cylinder.
Because a uniform analytic expansion of the solution in the whole domain is
not possible, we derived two different expansions in
the two subregions $\Omega_i$ and $\Omega_o$, and then matched them at the boundary between the two subcompartments.

We first analyzed the narrow escape time (NET) $\tau(r,z)$ to reach
the small hole in a laterally closed cylinder. { For a flat
cylinder with $h \ll R$ and $R\gg a$, we obtained that the NET
(\ref{meanexittime3_smallbeta}) is given by
\bea \label{newf}
\tau \approx \frac{|V|}{aD}\frac{a_0\(\frac{h}{a}\)}{\sqrt 2}  + \frac{R^2}{8D} \(4\ln\(\frac{R}{a}\) -3 \)\,,
\eea
where the function $a_0(\beta)\in [0.07,0.25]$ (Fig.~\ref{fig_a0}a).
Although (\ref{newf}) was derived in the condition that $h \ll R$, our
numerical results suggest that it remains valid until $h\sim R$
(section \ref{subsubsection_a0}). In particular, for a cylinder with
$h/a \gg  \ln\(\frac{R}{a}\)$ we recover the well known NET
approximation $\tau \approx \frac{|V|}{4aD}$
\cite{Ward1,Grigorievetal2002,HolcmanSchuss2004_StatPhys}, which was
derived for non degenerated geometries (isoperimetric ratio of order 1 and no bottle neck).
However, our analysis for the cylinder revealed that $\tau=\frac{|V|}{4aD}$ is
already a very good approximation for an oblate volume with an
isoperimetric ratio that can be very different from 1 (although $h/a \gg
\ln\(\frac{R}{a}\)$, the ratio $h/R$ can still be small).} Formula \ref{newf}
can also be used to estimate the rate constant of a key chemical reaction
during the early stage of phototransduction, which is the rate for
diffusing cGMP molecules to reach the phosphodiesterase enzyme
located on the surface of a narrow cylinder, located in the outer
segment of a rod photoreceptor \cite{ReingruberHolcman_JCP2008,ReingruberHolcman_PRE2009}.

In a next step, we used our method to analyze the narrow escape
problem for a flat and laterally open cylinder, which is
relevant for synaptic transmission. In many cases, the geometry of
the synaptic cleft is well approximated by a laterally open cylinder
\cite{Barbour_JNeurosc2001,Lisman_JNeurophys2004,TafliaHolcman2010},
where neurotransmitters are released into the synaptic cleft from
the presynaptic terminal, located on the upper surface. The
neurotransmitters move in the synaptic cleft
(Fig.~\ref{Domain_cylinder}) by Brownian diffusion, and they either activate receptors
clustered in the postsynaptic density located on the postsynaptic terminal (corresponding to the lower
cylinder surface), or they leave the synaptic cleft through
the lateral boundary without binding to a receptor.

We estimated the conditional probability $p(r,z)$ that a particle starting at position $(r,z)$
reaches the small hole before leaving the cylinder. By identifying
the small hole with the postsynaptic density where receptors are
clustered, we estimated the fraction of released neurotransmitters
that reach the receptor area before leaving the synaptic cleft.
Using our analytic solution, we studied the impact of the synaptic
cleft geometry as well as the location of neurotransmitter release.
In Fig.~\ref{fig_prob_pi(x,beta)_dim3}, we plotted the probability to
reach the postsynaptic density before leaving as a function of the
release position for various cylinder height, and we found that it is
very sensitive to the release position and the width of the
cylinder. We conclude that, in order to achieve an efficient
activation of postsynaptic receptors, the presynaptic and
postsynaptic densities should be properly aligned such that the
neurotransmitters are released opposite to the receptors. Finally, we computed the conditional mean time to reach the small hole before leaving through
the lateral boundary (see (\ref{expr_condMFPT})). For a wide cylinder, we found that the
conditional mean time is roughly by a factor $(2\ln \alpha)^2$ faster compared to the NET in a closed cylinder (see (\ref{approxtauctau})).

We shall now present some numerical estimates for neurotransmitters that need to activate receptors clustered in the postsynaptic density with a radius $a=50 nm$, when the synaptic cleft has a height $h=20 nm$ and a total radius of $R=500 nm$, so that $\alpha=R/a=10$ and $\beta=h/a=0.4$. When the transmitter are released at distance $r$ away from the center, the conditional probability is approximately given by truncating (\ref{solutionPi}) at $n=0$
\bea
p(r,h)=\left\{
\begin{array}{l}
\displaystyle  1 - b_0^p \frac{I_0( \frac{\pi r}{2h})}{I_0( \frac{a \pi}{2h})} \,, \quad x\le 1 \\ \\
\displaystyle  \(1 - \frac{a_0^p}{\sqrt 2} \)\frac{\ln \( \frac{R}{r}\) }{\ln R/a}
\,, \quad 1 \le r \le  R
\end{array}\right.
\eea
which is a very good approximation, as shown in Fig.~\ref{fig_prob_trunc}. Using a diffusion constant $D=200 \mu m^2/s$, we obtain for the mean times $\tau(0,h)\approx17 \mu s$ and $\tau_c(0,h)=1\mu s$.

The exact solution for the mean time and the conditional probability were
obtained here by using the patching eigenfunction expansion approach, and from
these approximations, we derived in the limit of large aspect ratio $R/h\gg1$ the asymptotic behavior.
Our approach works well because of the radial symmetry due to the absorbing trap that is is located at the center of the cylinder. This situation accounts well for the postsynaptic density located at the center of the post-synaptic terminal.  However, using our method it would be difficult to treat the case of multiple non-concentric traps, and in this case a different approach based on matching asymptotic analysis should be more appropriate   \cite{WardPillay_Siam2010,WardCheviakov_Siam2010}.
The analysis should start with an explicit representation of the Green's function for a cylinder. Once an inner solution is determined near each trap, it should be matched to the outer solution \cite{WardPillay_Siam2010}. This method should allow to study the effect of the trap positions and trap clustering on the synaptic current, which was only partially discussed in \cite{TafliaHolcman2010,HolcmanSchuss_PhysLettA2008}.

\subsubsection*{Acknowledgements}
D.H. research is supported by an ERC Starting Grant.

\cleardoublepage

\section*{APPENDIX}

\appendix

\section{Equation for the parameters $a_n$ in the limit  $\beta \ll 1$}
\label{asymptoticForBetaTo0}

In order to find the asymptotic equations for the coefficients $a_n$ for $\alpha \gg 1$ and $\beta \to 0$, we
introduce the scaled quantities
\bea
\begin{array}{c}
\ds \hat l_n = \beta l_n = \frac{(2n+1)\pi}{2}\,,\quad
\hat k_n = \beta k_n = n\pi \,,\\
\ds \hat \beta_n = \beta \beta_n = \hat l_n
\frac{I_1(l_n)}{I_0(l_n)}\,,\quad
\hat \alpha_n = \beta \alpha_n= \hat k_n \frac{K_1(k_n)}{K_0(k_n)} \,, \quad
\hat \gamma_0 = \beta \gamma_0 = \frac{1}{\sqrt 2 \pi}\,.\nn
\end{array}
\eea
In the limit $\beta \to 0$ we have $l_n\to \infty$ and $k_n\to \infty$ for every $n>0$, and
the asymptotic behaviour of $\hat \beta_n$ and $\hat \alpha_n$ is
\bea
\begin{array}{c}
\hat \beta_n \approx  \hat l_n \,, \quad  n\hat \alpha_n \approx  \hat k_n
\end{array} \nn
\eea
Using (\ref{eqanasymp}), the asymptotic equation for the coefficients $ a_n$ in the limit $\beta \to 0$ is
\bea\label{eqanbetato0}
\sum_{m=0}^\infty  (\hat l_n + \hat k_m ) \xi_{nm} a_m =\frac{1}{\sqrt 2 \pi} \xi_{n0}\,.
\eea
Truncating (\ref{eqanbetato0}) at various levels $n$ gives the  approximations
\bea
\frac{a_0}{\sqrt 2}  &=&  \frac{1}{\pi ^2} \approx 0.10 \,,\quad  n=0 \nn\\
\frac{a_0}{\sqrt 2} &=&   \frac{5}{6} \frac{1}{\pi^2}\approx 0.085\,,   \quad  n=1 \nn\\
\frac{a_0}{\sqrt 2} &=&    \frac{47}{60} \frac{1}{\pi^2} \approx 0.078\,,  \quad  n=2\,. \nn
\eea
Numerically we find from (\ref{eqanbetato0})
\bea
\frac{a_0}{\sqrt 2} \approx 0.071
\eea

\bibliographystyle{plain}

\end{document}